\documentclass[11pt]{article}
\usepackage{t1enc}
\usepackage{enumerate}
\usepackage{amssymb,amsmath,color,xcolor}
\usepackage{graphicx}
\usepackage{natbib}
\usepackage{xr-hyper}
\usepackage[citecolor = blue,linkcolor = orange, filecolor = orange]{hyperref}   
\hypersetup{colorlinks=true}
\usepackage[draft,multiuser]{fixme}
\fxusetheme{colorsig}
\FXRegisterAuthor{peter}{Peter}{\color{blue}P} 
\FXRegisterAuthor{teemu}{Teemu}{\color{red}R}
\DeclareRobustCommand{\VANDER}[3]{#2}

\newtheorem{examplehidden}{Example}
\newenvironment{example}{
\begin{examplehidden}
\em
} 
{$\Box$\end{examplehidden}}

\newcommand{\ud}[2]{\bar{p}^{#1}_{#2}}
\newcommand{\pud}[3]{{p}^{#1}_{{#2}_{#3}}}
\newcommand{\full}{\text{\sc full}}
\newcommand{\mdlest}[2]{\hat{#1}_{#2}}
\newcommand{\mlest}[1]{{\hat{#1}_{\text{\sc ml}}}}
\newcommand{\mlestgamma}[2]{{\hat{#1}_{\text{\sc ml}| #2}}}
\newcommand{\jp}{w_{\text{\sc J}}}
\newcommand{\comp}{\text{\sc comp}}
\newcommand{\cM}{{\cal M}}
\newcommand{\cZ}{{\cal Z}}

\newcommand{\cY}{{\cal Y}}

\newcommand{\reals}{\mathbb{R}}

\newcommand{\bayes}{\text{\sc bayes}}
\newcommand{\nml}{\text{\sc nml}}
\newcommand{\preq}{\text{\sc preq}}
\newcommand{\twopart}{\text{\sc 2-p}}
\newcommand{\switch}{\text{\sc switch}}
\newcommand{\ripr}{\text{\sc ripr}}
\newcommand{\regret}{\text{\sc REG}}

\newcommand{\normal}{\text{\sc gauss}}

\title{Minimum Description Length Revisited}
\author{Peter Gr\"unwald\thanks{Also affiliated with Leiden University, The Netherlands.} \\ CWI \\ 
P.O. Box 94079, 1090 GB Amsterdam The Netherlands \and Teemu Roos\footnote{Also affiliated with HIIT (Helsinki Insitute of Information Technology), Helsinki, Finland.} \\ University of Helsinki \\ P.O. Box 68, FI-000014 Finland}

\begin{document}
\maketitle
\abstract{This is an up-to-date introduction to and overview of the Minimum Description Length (MDL) Principle, a theory of inductive inference that can be applied to general problems in statistics, machine learning and pattern recognition. While MDL was originally based on data compression ideas, this introduction can be read without any knowledge thereof. It takes into account all major developments since 2007, the last time an extensive overview was written. These include new methods for model selection and averaging and hypothesis testing, as well as the first completely  general definition of {\em MDL estimators}. Incorporating these developments, MDL can be seen as a powerful extension of both penalized likelihood and Bayesian approaches, in which
penalization functions and prior distributions are replaced by more general luckiness functions, average-case methodology is replaced by a more robust worst-case approach, and in which methods classically viewed as highly distinct, such as AIC vs BIC and cross-validation vs Bayes can, to a large extent, be viewed from a unified perspective. }
\section{Introduction}
The {\em Minimum Description Length (MDL) Principle\/} \citep{Rissanen78,Rissanen89,BarronRY98,Grunwald07} is a theory of inductive inference that can be applied to general problems in statistics, machine learning and pattern recognition. Broadly speaking, it states that the best explanation for a given set of data is provided by the shortest description of that data. In 2007, one of us published the  book {\em  The Minimum Description Length Principle\/} (\cite{Grunwald07}, G07
from now on), giving a detailed account of most work in the MDL area
that had been done until then. During the last 10 years, several new practical  MDL methods have been designed, and there have been exciting theoretical developments as well. It therefore seemed time to present an up-to-date combined introduction and review. 
\paragraph{Why Read this Overview?} While the MDL idea has been shown to be very powerful in theory, and there have been a fair number of successful practical implementations, massive deployment has been hindered by two issues: first, in order to apply MDL, one needs to have basic knowledge of both statistics and information theory. To remedy this situation, here we present, for the first time, the MDL Principle {\em without resorting to information theory\/}: all the material can be understood without any knowledge of data compression, which should make it a much easier read for statisticians and machine learning researchers novel to MDL. A second issue is that many classical MDL procedures are either computationally highly intensive (for example, MDL variable selection as in Example~\ref{ex:varsel} below) and hence less suited for our big data age, or they  seem to require somewhat arbitrary restrictions of parameter spaces (e.g. NML with $v\equiv 1$ as in Section~\ref{sec:universal}). Yet, over the last 10 years, there have been  exciting developments --- some of them very recent --- which mostly resolve these issues. Incorporating these developments, MDL  can be seen as a powerful extension of both penalized likelihood and Bayesian approaches, in which penalization functions and prior distributions are replaced by more general luckiness functions, average-case methodology is replaced by a more robust worst-case approach, and in which methods classically viewed as highly distinct, such as AIC vs BIC and cross-validation vs Bayes can, to some extent, be viewed from a unified perspective; as such, this paper should also be of interest to researchers working on the foundations of statistics and machine learning.  
\paragraph{History of the Field, Recent Advances and Overview of this Paper}
MDL was introduced in 1978 by Jorma Rissanen in his  paper {\em Modeling by the Shortest Data Description}. The paper coined the term {\em  MDL\/} and introduced and analyzed the two-part code for parametric models. The two-part code is the simplest instance of a {\em universal code\/} or, equivalently, {\em universal probability distribution}, the cornerstone concept of MDL theory. MDL theory was greatly extended in the 1980s, when Rissanen published a sequence of ground-breaking papers at a remarkable pace, several of which introduced new types of universal distributions. It came to full blossom in the 1990s, with further major contributions from, primarily, Jorma Rissanen, Andrew Barron and Bin Yu, culminating in their overview paper \citep{BarronRY98} and the collection \cite{grunwald2005advances} with additional chapters by other essential contributors such as Kenji Yamanishi. The book G07 provides a more exhaustive treatment of this early work, including discussion of important precursors/alternatives to MDL such as MML \citep{WallaceB68}, `ideal', Kolmogorov complexity based MDL \citep{VitanyiL00} and Solomonoff's theory of induction \citep{sterkenburg2018universal}. Universal distributions are still central to MDL. We introduce them in a concise yet self-contained way, including substantial underlying motivation, in Section~\ref{sec:universal}, incorporating the extensions to and new insights into these basic building blocks that have been gathered over the last 10 years. These include more general formulations of arguably the most fundamental universal code, the {\em Normalized Maximum Likelihood (NML) Distribution}, including faster ways to calculate it as well. We devote a separate section to new universal codes, with quite pleasant properties for practical use, most notably the {\em switch
  distribution\/} (Section~\ref{sec:switch}), which can be used for model selection combining almost the best of AIC and BIC; and the {\em RIPr}-universal code (Section~\ref{sec:svalues}) specially geared to hypothesis testing with composite null hypotheses, leading to several advantages over classical Neyman-Pearson tests.
In Section~\ref{sec:graphical} we review recent developments on fast calculation of NML-type distributions for model selection for {\em graphical models\/} (Bayesian networks and the like), leading to methods which appear to be more robust in practice than the standard Bayesian ones.
 Recent extensions of MDL theory and practical implementations to latent variable and irregular models are treated in Section~\ref{sec:latent}. 
Then, in Section~\ref{sec:freq} we review developments relating to consistency and convergence properties of MDL methods.   
First, while originally MDL estimation was formulated solely in terms of discretized estimators (reflecting the fact that coding always requires discretization), it has gradually become clear that a much larger class of estimators (including maximum likelihood for `simple' models, and, in some circumstances, the Lasso --- see Example~\ref{ex:varsel}) can be viewed from an MDL perspective, and this becomes clearest if one investigates asymptotic convergence theorems relating to MDL. Second, it was found that MDL (and Bayes), without modification, can behave
sub-optimally under misspecification, i.e. when all models under consideration are wrong, but some are useful --- see Section~\ref{sec:misspecification}. Third, very
recently, it was shown how some of the surprising phenomena underlying the {\em deep learning\/} revolution in machine learning can be explained from an MDL-related perspective; we briefly review these developments in Section~\ref{sec:deep}. 
Finally, we note that G07 presented many explicit open problems, most of which have been resolved --- we mention throughout the text whenever a new development solved an old open problem, deferring some of the most technical issues to Appendix~\ref{app:nml}.
\paragraph{Notational Preliminaries} 
We shall mainly be concerned with {\em statistical models\/} (families of probability distributions) of the form $\cM = \{ p_{\theta}: \theta \in \Theta \}$ parameterized by some set $\Theta$ which is usually but not always a subset of Euclidean space; and {\em families of models\/} $\{{\cal M}_{\gamma} : \gamma \in \Gamma \}$, where each ${\cal M}_{\gamma} = \{p_{\theta}: \theta \in \Theta_{\gamma} \}$ is a statistical model, used to model data $z^n := (z_1, \ldots, z_n)$ with each $z_i \in \cZ$, for some outcome space $\cZ$. 
Each $p_{\theta}$ represents a probability density function (pdf) or probability mass function, defined on sequences of arbitrary length. With slight abuse of notation we also denote the corresponding probability distribution by $p_{\theta}$ (rather than the more common $P_{\theta}$). In the simple case that the data are i.i.d. according to each $p_{\theta}$ under consideration, we have  $p_{\theta}(z^n) = \prod_{i=1}^n p_{\theta}(z_i)$. 

We denote the ML (maximum likelihood) estimator given model ${\cal M}
= \{p_{\theta} : \theta \in \Theta \}$ by $\mlest{\theta}$, whenever it
exists and is unique; the ML estimator relative to model $\cM_{\gamma}$ is denoted $\mlestgamma{\theta}{\gamma}$. We shall, purely for
simplicity, generally assume its existence and uniqueness, although nearly all results can be generalized to the case where it does not. We use
$\breve\theta$ to denote more general estimators, and $\mdlest{\theta}{v}$ to denote what we call the {\em MDL estimator with luckiness function $v$}, see (\ref{eq:brevepre}). 
 
\section{The Fundamental Concept: Universal Modeling}
\label{sec:universal}
MDL is best explained by starting with one of
its prime applications, model comparison --- we will generalize to
prediction and estimation later, in
Section~\ref{sec:large} and \ref{sec:codelength}. Assume then that we are given a finite
or countably infinite collection of statistical models $\cM_1, \cM_2, \ldots$, each consisting of a
set of probability distributions.  The fundamental idea of MDL is to
associate each $\cM_{\gamma}$ with a {\em single\/} distribution
$\ud{}{\gamma}$, often called a {\em universal distribution\/} relative
to $\cM_{\gamma}$. We call the minus-log-likelihood $-\log \ud{}{\gamma}(Z^n)$ the {\em code length of data $Z^n$ under universal code
  $\ud{}{\gamma}$}. This terminology, and how MDL is
related to coding (lossless compression of data), is briefly reviewed in
Section~\ref{sec:large} and Section~\ref{sec:codelength}; but a crucial observation at this point is that the main MDL ideas can be understood abstractly, without resorting to the code length interpretation.  We also equip the model indices $\Gamma:= \{1,2,
\ldots, \gamma_{\max} \}$ (where we allow $|\Gamma| = \gamma_{\max} = \infty$) with a
distribution, say $\pi$; if the number of models to be compared is
small (e.g. bounded independently of $n$ or at most a small polynomial in $n$), we can take $\pi$ to be uniform distribution --- for large (exponential in $n$) and
infinite $\Gamma$, see Section~\ref{sec:large} and Example~\ref{ex:varsel}. 
We then take, as our best explanation of the given data $z^n$, the
 model $\cM_{\gamma}$ minimizing
\begin{equation}\label{eq:selectingk}
- \log \pi(\gamma) - \log \ud{}{\gamma}(z^n ),
\end{equation}
or, equivalently, we maximize $\pi(\gamma) \ud{}{\gamma}(z^n)$; when $\pi$ is uniform this simply amounts to picking the $\gamma$ maximizing $\ud{}{\gamma}(z^n)$. (\ref{eq:selectingk}) will later be generalized to $\pi$ that are not distributions but rather more general `luckiness functions' --- see Section~\ref{sec:large}.

\paragraph{1. The Bayesian Universal Distribution}
The reader may recognize this as being formally equivalent
to the standard Bayesian way of model selection, the {\em Bayes factor
  method\/} \citep{kass1995bayes} as long as the
${}{\gamma}$ are defined as Bayesian marginal
distributions, i.e. for each $\gamma$, we set $\ud{}{\gamma} = \pud{\bayes}{w}{\gamma}$, where
\begin{equation}\label{eq:bayesuniversal}
\pud{\bayes}{w}{\gamma}(z^n):= \int p_{\theta}(z^n) w_{\gamma}(\theta) d\theta,
\end{equation}
for some prior probability density $w_{\gamma}$
on the parameters in $\Theta_{\gamma}$, which has to be supplied by the
user. When $w_{\gamma}$ is clear from the context, we shall write $\ud{\bayes}{\gamma}$ rather than $\pud{\bayes}{w}{\gamma}$. Using Bayesian marginal distributions $\ud{\bayes}{}$ is indeed one possible way to instantiate MDL model selection, but it is not the only way: MDL can also be based on other distributions such as $\ud{\nml}{} = \pud{\nml}{v}{}$ (depending on a function $v$) , $\ud{\preq}{} = \pud{\preq}{\breve\theta}{}$ (depending on an estimator $\breve\theta$)  and others; in general we add a bar to such distributions if the `parameter' $w,v$ or $\breve\theta$ is clear from the context. Before we continue with these other instantiations of $\ud{}{\gamma}$ we proceed with an example: 
\begin{example}
\label{ex:bernoulli}{\bf \ [Bernoulli]}
Let  $\cM = \{p_{\theta}: \theta \in [0,1]\}$ represent the Bernoulli model, extended to $n$ outcomes by independence. We then have for each $z^n \in \{0,1\}^n$ that $p_{\theta}(z^n) = \theta^{n_1} (1- \theta)^{n_0}$,
where $n_1 = \sum_{i=1}^n z_i$ and $n_0 = n- n_1$. Most standard prior distributions one encounters in the literature are beta priors, for which $w(\theta) \propto \theta^{\alpha} (1- \theta)^{\beta}$, so that $\pud{\bayes}{w}{}(z^n) \propto \int \theta^{n_1 + \alpha}(1-\theta)^{n_0 + \beta} d \theta$. Note that $\pud{\bayes}{w}{}$ is not itself an element of the Bernoulli model. One could use $\pud{\bayes}{w}{}$ to compare the Bernoulli model, via (\ref{eq:selectingk}), to, for example, a first order Markov model, with Bayesian marginal likelihoods defined analogously. We shall say a lot more about the choice of prior below.
\end{example}
\begin{example}\label{ex:gauss}{\bf \ [Gauss and general improper priors]} A second example is the Gaussian location family $\cM_{\normal}$ with fixed variance (say 1), in which $\cZ = {\mathbb R}$, and $p_{\theta}(z^n) \propto \exp(\sum_{i=1}^n (z_i - \theta)^2/2)$. A standard prior for such a model is the uniform prior, $w(\theta) = 1$, which is {\em improper\/} (it does not integrate, hence does not define a probability distribution). Improper priors cannot be directly used in (\ref{eq:bayesuniversal}), and hence they cannot be directly used for model comparison as in (\ref{eq:selectingk}) either. Still, we can use them in an indirect manner, as long as we are guaranteed that, for all $\cM_{\gamma}$ under consideration, after some initial number of $m$ observations, the Bayesian posterior $w_{\gamma}(\theta \mid z^m)$ is proper. We can
then
replace $\pud{\bayes}{w}{{\gamma}}(z^n)$ in (\ref{eq:bayesuniversal}) by $\pud{\bayes}{w}{{\gamma}}(z_{m+1}, \ldots, z_n \mid z^m) := \int p_{\theta}(z_{m+1}, \ldots, z_n) w_{\gamma}(\theta \mid z^m) d \theta$.
We extend all these {\em conditional universal distributions\/} to distributions on $\cZ^n$ by defining $\pud{\bayes}{w}{\gamma}(z_1, \ldots, z_n) := \pud{\bayes}{w}{\gamma}(z_{m+1}, \ldots, z_n \mid z^m) p_0(z^m)$ for some distribution $p_0$ on $\cZ^m$ that is taken to be the same for all models $\cM_{\gamma}$ under consideration. We can now use (\ref{eq:selectingk}) again for model selection based on the $\pud{\bayes}{w}{{\gamma}}(z_1, \ldots, z_n)$, where we note that the choice of $p_0$ plays no role in the minimization, which is equivalent to minimizing $- \log \pi(\gamma) - \log\pud{\bayes}{w}{{\gamma}}(z_{m+1}, \ldots, z_n \mid z^m)$. 
\end{example}
Now comes the crux of the story, which makes MDL, in the end, quite different from Bayes: defining the $\ud{}{\gamma}$ as in (\ref{eq:bayesuniversal}) is
just {\em one particular way\/} to define an MDL universal
distribution --- but it is by no means the only one. There are several
other ways, and some of them are sometimes preferable to the Bayesian choice. Here we list the
most important ones:
\paragraph{2. NML (Normalized Maximum Likelihood) or
  {\em Shtarkov\/} [\citeyear{Shtarkov87}] distribution, and MDL estimators} 
  This is perhaps the most fundamental universal distribution, leading also to the definition of an {\em MDL estimator}.
  In its general form, the NML distribution and `MDL estimators' depend on a function $v: \Theta
  \rightarrow \reals^+_0$. The definition is then given by
\begin{equation}\label{eq:nmluniversal}
\pud{\nml}{v}{}(z^n):= 
\frac{\max_{\theta \in \Theta} p_{\theta}(z^n) v(\theta)}{
\int \max_{\theta \in \Theta} p_{\theta}(z^n) v(\theta) d z^n
} \overset{\text{(if $v$ constant)}}{=}  
\frac{p_{\mlest{\theta}(z^n)}(z^n) }{
\int p_{\mlest{\theta}(z^n)}(z^n) d z^n
},
\end{equation}
which is defined whenever the normalizing integral is finite. The logarithm of this integral is called the {\em model complexity} and is thus given by 
\begin{equation}\label{eq:complexitypre}
    \comp(\cM; v) := \log \int \max_{\theta \in \Theta} (p_{\theta}(z^n) v(\theta)) d z^n \overset{\text{(if $v$ constant)}}{=}  
\log {
\int p_{\mlest{\theta}(z^n)}(z^n) d z^n
}.\end{equation}
Here the integral is replaced by a sum
for discrete data, and $\max$ is replaced by $\sup$ if necessary. This means that any function $v: \Theta \rightarrow \reals^+_0$ such that (\ref{eq:complexitypre}) is finite is allowed; we call any such $v$ a {\em luckiness function}, a terminology we explain later. 
Note that $v$ is not necessarily a probability density --- it does not have to be integrable. For any luckiness function $v$, we define the {\em MDL estimator based on $v$\/} as
\begin{equation}\label{eq:brevepre}
\mdlest{\theta}{v} := \arg \max_{\theta \in \Theta} p_{\theta}(z^n) v(\theta) = \arg \min_{\theta \in \Theta}  \ \{\ - \log p_{\theta} - [- \log v(\theta)] \ \}
\end{equation}
The $v$-MDL estimator is  a penalized ML estimator, which coincides 
with the  Bayes MAP estimator based on prior $v$  whenever $v$ is a probability density.
Although this has only become clear gradually over the last 10 years, estimators of form (\ref{eq:brevepre}) are the prime way of using MDL for estimation; there is, however, a second, `improper' way for estimating distributions within MDL though, see Section~\ref{sec:codelength}.  In practice, we will choose $v$ that are sufficiently smooth so that, if the number of parameters is small relative to $n$, $\mdlest{\theta}{v}$ will usually be almost indistinguishable from the ML estimator $\mlest{\theta}$. $\comp$ indeed measures something one could call a `complexity' --- this is easiest to see if $v \equiv 1$, for then, if $\cM$ contains just a single distribution, we must have $\comp(\cM,v) = 0$, and the more distributions we add to $\cM$, the larger $\comp(\cM,v)$ gets --- this is explored further in Section~\ref{sec:asymptotic}.

Now suppose we have a collection of models $\cM_{\gamma}$ indexed by finite $\Gamma$ and we have specified luckiness functions $v_{\gamma}$ on $\Theta_\gamma$ for each $\gamma \in \Gamma$, and we pick a uniform distribution $\pi$ on $\Gamma$. As can be seen from the above, if we base our model choice on NML, we pick the model minimizing 
\begin{equation}\label{eq:tradeoff}
    - \log p_{\mdlest{\theta}{v_\gamma}(z^n)}(z^n) - \log v_{\gamma}(\mdlest{\theta}{v_\gamma}(z^n)) \ + \comp(\cM_{\gamma}; v_{\gamma}),
\end{equation}
over $\gamma$, where $\comp(\cM_{\gamma}; v_{\gamma})$ is given by
\begin{equation}\label{eq:complexity}
    \comp(\cM_{\gamma}; v_{\gamma}) = \log \int \max_{\theta \in \Theta_{\gamma}} (p_{\theta}(z^n) v_{\gamma}(\theta)) d z^n = 
\log {
\int p_{\mdlest{\theta}{v_\gamma}(z^n)}(z^n) v_{\gamma}(\mdlest{\theta}{v_\gamma}(z^n))  d z^n
}.\end{equation}
Thus, by (\ref{eq:tradeoff}), MDL incorporates a trade-off between goodness-of-fit and model complexity as measured by $\comp$. Although the $n$-fold integral inside $\comp$ looks daunting, \cite{suzuki2018exact} show that in many cases (e.g. normal, Weibull, Laplace models) it can be evaluated explicitly with appropriate choice of $v$.

Originally, the NML distribution was defined by \cite{Shtarkov87} for the special case with $v \equiv 1$, leading to the rightmost definition in (\ref{eq:nmluniversal}), and hence the term NML (in the modern version, perhaps `normalized {\em penalized\/} ML' would be more apt). This
is also the version that \cite{Rissanen96} advocated as embodying the
purest form of the MDL Principle. However, the integral in
(\ref{eq:nmluniversal}) is ill-defined for just about every parametric
model defined on unbounded outcome spaces (such as ${\mathbb N}, {\mathbb R}$ or ${\mathbb R}^+$), including the simple
normal location family. Using nonuniform $v$ allows one to deal with such cases in a principled manner after all,
see Section~\ref{sec:luckiness}.
For finite outcome spaces though, $v \equiv 1$ usually `works', and (\ref{eq:nmluniversal}) is well defined, as we illustrate for the Bernoulli model (see Section~\ref{sec:graphical} for more examples):
\begin{example}{\bf [Example~\ref{ex:bernoulli}, Cont.]}
For the Bernoulli model, $\mlest{\theta}(z^n) = n_1/n$ and $\comp(\cM,v)$ as in (\ref{eq:complexity}) with $v \equiv 1$ can be rewritten as $\log  \sum_{n_1 = 0}^{n} \binom{n}{n_1} (n_1/n)^{n_1} (n_0/n)^{n_0}$, which, as we shall see in Section~\ref{sec:asymptotic}, is within a constant of $(1/2) \log n$. As reviewed in that section, the resulting $\pud{\nml}{v}{}$ is asymptotically (essentially) indistinguishable from $\pud{\bayes}{\jp}{}$ where the latter is equipped with {\em Jeffreys' prior}, defined as  $\jp(\theta) \propto \sqrt{|I(\theta)} = \theta^{-1/2} (1-\theta)^{-1/2}$, 
$I(\theta)$ being the Fisher information at $\theta$.
\end{example}
 \paragraph{3. The  two-part (sub-) distribution} \citep{Rissanen78}. 
Here one first discretizes $\Theta$ to some countable subset $\ddot{\Theta}$ which one equips with a probability mass function $w$; in contrast to the $v$ above,  this function must sum to $1$.  One then considers 
\begin{equation}\label{eq:twopart}
    \pud{\nml}{w}{}(z^n) := \frac{\max_{\ddot{\theta} \in \ddot{\Theta}} p_{\ddot{\theta}}(z^n) w(\ddot{\theta})}{
    \int \max_{\ddot{\theta} \in \ddot{\Theta}} p_{\ddot{\theta}}(z^n) w(\ddot{\theta}) d z^n},
\end{equation}
which is just a special case of (\ref{eq:nmluniversal}). But since
\begin{equation}\label{eq:twopartone}
\int \max_{\ddot{\theta} \in \ddot{\Theta}} p_{\ddot{\theta}}(z^n) w(\ddot{\theta}) d z^n \leq \int \sum_{\ddot{\theta} \in \ddot{\Theta}} p_{\ddot{\theta}}(z^n) w(\ddot{\theta}) d z^n  = 
\sum_{{\theta} \in \ddot{\Theta}} w({\theta}) \left(  \int p_{{\theta}}(z^n) d z^n \right) =1,
\end{equation}
we can approximate $\pud{\nml}{w}{}$ by the {\em sub}-distribution $\pud{\twopart}{w}{}(z^n) := \max_{\ddot{\theta} \in \ddot{\Theta}} p_{\ddot{\theta}}(z^n) w(\ddot{\theta})$. This `distribution'  adds or integrates to something smaller than $1$. This can be incorporated into the general story by imagining that $\pud{\twopart}{w}{}$ puts its remaining mass on a special outcome, say `$\diamond$',
which in reality will never occur (while sub-distributions are thus `allowed', measures that add up to something {\em larger\/} than 1 have no place in MDL). The two-part distribution $\pud{\twopart}{w}{}$ is historically the oldest universal distribution. The fact that it can be considered a special case of NML has only become fully clear very recently  \citep{GrunwaldM19}; in that same paper, an even more general formulation of (\ref{eq:nmluniversal}) is given that has all Bayesian, two-part and NML distributions as special cases. Despite its age, the two-part code is still important in practice, as we explain  in Section~\ref{sec:large}.
\paragraph{4. The prequential plug-in distribution} \citep{Rissanen84,Dawid84}. Here, one first takes any reasonable estimator $\breve\theta$ for  the given model $\cM$. 
One then defines
\begin{equation}\label{eq:prequential}
\pud{\preq}{\breve\theta}{}(z^n) := \prod_{i=1}^n p_{\breve\theta(z^{i-1})}(z_i \mid z^{i-1}),
\end{equation}
where for i.i.d. models, the probability inside the product simplifies to
$p_{\breve\theta(z^{i-1})}(z_i)$. For the normal location family, one could simply use the ML estimator: $\breve\theta(z^{m}) := \mlest{\theta}(z^{m}) = \sum_{j=1}^{m} z_j/m$. With discrete data though, the ML estimator should be avoided, since then one of the factors in (\ref{eq:prequential}) could easily become 0, making the  product 0, so that the model for which $\pud{\preq}{\breve{\theta}}{}$ is defined can never `win' the model selection contest even if most other factors in the product (\ref{eq:prequential}) are close to 1. Instead, one can use a slightly  `smoothed' ML estimate (a natural choice for $\breve{\theta}$ is to take an MDL estimator for some $v$ as in (\ref{eq:brevepre}), but this is not required). For example, in the Bernoulli model, one might take $\breve\theta(z^m) = (m_1 +(1/2))/(m+1)$, where $m_1 = \sum_{i=1}^m z_i$. With this particular choice, $\pud{\preq}{\breve\theta}{}$ turns out to coincide {\em exactly\/} with $\pud{\bayes}{w}{{\text{\sc J}}}$ with Jeffreys' prior $w_{\text{\sc J}}$. Such a precise correspondence between $\ud{\preq}{}$ and $\ud{\bayes}{}$ is a special property of the Bernoulli and multinomial models though; with other models, the two distributions can usually be made to behave similarly, but not identically. The rationale for using $\ud{\preq}{}$ is described in Section~\ref{sec:codelength}. In Section~\ref{sec:kotlowski} we will say a
bit more about hybrids between prequential plug-in and Bayes (the {\em flattened leader distribution\/}) and between prequential and NML ({\em sequential
  NML}).
\\
\ \newline
Except for the just mentioned `hybrids', these first four universal distributions were all brought into MDL theory by Rissanen; they are extensively treated by G07, who devoted one chapter to each, and to which we refer for details.  The following two are much more recent:
\paragraph{5. The  switch distribution $\ud{\switch}{}$} \citep{ErvenGR07}. In a particular type of nested model selection, this universal distribution behaves arguably better than the other ones. It will
  be treated in detail in Section~\ref{sec:switch}.
\paragraph{6. Universal distributions $\ud{\ripr}{}$ based on the Reverse Information Projection (RIPr)} These universal distributions \citep{GrunwaldHK19} lead to improved error bounds and optional stopping behaviour in hypothesis testing and allow one to forge a connection with group invariant Bayes factor methods; see Section~\ref{sec:svalues}. 
\subsection{Motivation}\label{sec:motivation}
We first give a very high-level motivation that avoids direct use of data compression arguments. For readers interested in data compression, Section~\ref{sec:large} does make a  high-level connection, but for more extensive material we refer to G07.  We do, in Section~\ref{sec:codelength}, give a more detailed motivation in predictive terms, and, in Section~\ref{sec:freq}, we shall review mathematical results indicating that MDL methods are typically consistent  and enjoy fast  rates of convergence, providing an additional motivation in itself. 

Consider then models $\cM_\gamma$, where for simplicity we assume discrete data, and let $\mlestgamma{\theta}{\gamma}$ be the
maximum likelihood estimator within $\cM_{\gamma}$. Define `the fit of the model to the
data' in the standard way, as $F_{\gamma}(z^n): = p_{\mlestgamma{\theta}{\gamma}(z^n)}(z^n)$,
the likelihood assigned to the data by the best-fitting distribution
within model. Now if we enlarge the model $\cM_{\gamma}$, i.e. by adding several
distributions to it, $F_{\gamma}(z^n)$ can only increase; and if we make $\cM_{\gamma}$ big enough such that for each $z^n$, it contains a distribution $p$
with $p(z^n) = 1$, we can even  have $F_{\gamma}(z^n) = 1$ on all data. If we simply picked the $\gamma$ maximizing $F_{\gamma}(z^n)$, we would be prone to severe overfitting. For example, if models are nested, then, except for very special data, we would automatically pick the largest one.

As we have seen, a central MDL idea is to instead
associate each model $\cM_{\gamma}$ with a {\em single\/}
corresponding distribution $\ud{}{\gamma}$, i.e. we set  $F_{\gamma}(z^n) := \ud{}{\gamma}(z^n)$. Then the total probability mass on all potential outcomes $z^n$ cannot be larger than
$1$, which makes it impossible to assign overly high fit $F_{\gamma}(z^n)$ to
overly many data sequences: no matter what distribution $\ud{}{\gamma}$ we
chose, we must now have $\sum_{z^n} F_{\gamma}(z^n) = 1$, so a good fit on some $z^n$ necessarily implies a worse fit on others, and we will not select a model simply because it accidentally contained {\em some\/} distribution that fitted our data very well -- thus, measuring fit by a distribution $\ud{}{\gamma}$ instead of $F_{\gamma}$ inherently prevents overfitting. This argument to measure fit relative to a model with a single $\ud{}{}$ is
similar to {\em Bayesian Occam's Razor\/} arguments
\citep{RasmussenG00} used to motivate the Bayes factor; the crucial
difference is that we do not restrict ourselves to $\ud{}{\gamma}$ of the form (\ref{eq:bayesuniversal}); inspecting the ``Bayesian'' Occam argument, there is, indeed nothing in there which forces us to use distributions of Bayesian form. 

The next step is thus to decide {\em which\/} $\ud{}{}$ are best associated with a given 
$\cM$. 
To this end, we define the {\em fitness ratio\/} for data $z^n$ as
\begin{equation}\label{eq:expregret}
\text{\sc FR}(\ud{}{},z^n) := \frac{\ud{}{}(z^n)}{
\max_{\theta \in \Theta} p_{\theta}(z^n) v (\theta)}, 
\end{equation}
where $v: \Theta \rightarrow {\mathbb R}^+_0$ is a nonnegative function.  To get a feeling for (\ref{eq:expregret}), it is best to first focus on the case with $v(\theta) \equiv 1$; it then reduces to
\begin{equation}\label{eq:expregretb}
\text{\sc FR}(\ud{}{},z^n) = \frac{\ud{}{}(z^n)}{
p_{\mlest{\theta}(z^n)}(z^n)}.
\end{equation}
We next postulate that a good choice for $\ud{}{}$ relative to the
given model is one in which $\text{\sc FR}(\ud{}{},n)$ {\em tends to
  be as large as possible}. The rationale is that, overfitting having already been taken care of by picking some $\ud{}{}$ that is a probability measure (integrates to 1), it  makes
sense to take a $\ud{}{}$ whose fit to data (as measured in terms of
likelihood) is proportional to the fit to data of the best-fitting
distribution in $\cM$:  whenever {\em some\/} distribution in the model $\cM$ fits the data $z^n$ well, the likelihood $\ud{}{}(z^n)$ should be high as well.  
One way to make `$\text{\sc FR}$ tends to be large'  precise is by
requiring it to be as large as possible in the worst-case, i.e., we
want to pick the $\ud{}{}$ achieving
\begin{equation}\label{eq:maximin}
  \max_{\ud{}{}} \min_{z^n \in \cZ^n} \text{\sc FR}(\ud{}{},z^n),\end{equation}
where the maximum is over all probability distributions over samples
of length $n$. It turns out that this maximin problem has a solution if and only if the complexity (\ref{eq:complexitypre}) is finite; and if it is fine, the unique solution is given by setting $\ud{}{} = \ud{\nml}{}$, with $\ud{\nml}{}$ given by (\ref{eq:nmluniversal}). The NML distribution thus has a special status as the most robust choice of universal $\ud{}{}$ --- even though $\ud{}{}$ is itself a probability distribution, it meaningfully assesses fit in the worst-case over all possible distributions, and its interpretation does not require one to assume that the model $\cM$ is `true' in any sense. 
The nicest sub-case is the one with $v(\theta) \equiv 1$, since then all distributions within the model $\cM$ are treated on exactly the same footing; no data or distribution is intrinsically preferred over any other one. 
Unfortunately, for most popular models with infinite $\cZ$, when taking $v(\theta) \equiv 1$,  (\ref{eq:maximin}) usually has no solution since the integral $\int p_{\mlest{\theta}(z^n)}(z^n) dz^n$ diverges for such models, making the complexity (\ref{eq:complexitypre}) infinite. For all sufficiently `regular'  models (curved exponential families, see below), this problem can invariably be solved by restricting $\Theta$ to a bounded subset of its own interior - one can show that the complexity (\ref{eq:complexitypre}) is finite with $v \equiv 1$, and thus (\ref{eq:maximin}) has a solution given by (\ref{eq:nmluniversal}) if $\mlest{\theta}$ is restricted to a suitably bounded set. Yet, restricting $\Theta$ to a bounded subset of itself is not satisfactory, since it is unclear where exactly to put the boundaries. It is more natural to introduce a nonuniform $v$, which can invariably be chosen so that the complexity (\ref{eq:complexitypre}) is finite and thus (\ref{eq:maximin}) has a solution --- more on choosing $v$ at the end of Section~\ref{sec:codelength}. 

A few remarks  concerning this high-level motivation of MDL procedures are in order: 
\begin{enumerate}
    \item 
It is clear that, by requiring $\text{\sc FR}$  to add to 1, we will be less prone to overfitting than by setting it simply to $p_{\mlest{\theta}(z^n)}(z^n)$; whether the requirement to add (at most) to 1, making $\text{\sc FR}$ essentially a probability density function, is a {\em clever\/} way to avoid overfitting (leading to good results in practice) is not clear yet. For this, we need additional arguments, which we very briefly review. First, the sum-to-1 requirement is the only choice for which the procedure can be interpreted as selecting the model which minimizes code length of the data (the original interpretation of MDL); second, it is the only choice which has a predictive interpretation, which we review in Section~\ref{sec:codelength} below; third, it is the only choice under which time-tested Bayesian methods fit into the picture, and fourth, with this choice we get desirable frequentist statistical properties such as consistency and convergence rates, see Section~\ref{sec:freq}. 
\item The motivation above only applies to the NML universal distributions. How about the other five types? Originally, in the pure MDL approach mainly due to Rissanen, the NML was viewed as the optimal choice per se; other $\ud{}{}$ should be used only for pragmatic reasons, such as them being easier to calculate. One would then design them so as to be as close as possible to the NML distributions in terms of the fitness ratio they achieve. In the following subsection we show that all six of them satisfy the same MDL/BIC asymptotics, meaning that their fitness ratio is never smaller than a constant factor of the NML one, either again in the worst-case over all $z^n$ or in some weaker expectation sense. Thus, they are all `kind of o.k.' in a rather weak sense, and in practice one would simply revert to the one that is closest to NML and still usable in practice; with the Bayesian $\ud{\bayes}{}$, as we shall see, one can even get arbitrarily close to NML as $n$ gets larger. This classical story notwithstanding, it has become more and more apparent that in practice one sometimes wants or needs properties of model selection methods that are not guaranteed by NML -- such as near-optimal predictions of future data or strong frequentist Type-I error guarantees. This translates itself into universal codes $\ud{\switch}{}$ and $\ud{\ripr}{}$ that,  for some special sequences achieve much higher fitness ratio than $\ud{\nml}{}$, while for all sequences having only very slightly smaller fitness ratio. This more recent and pragmatic way of MDL is briefly reviewed in Section~\ref{sec:switch} and~\ref{sec:svalues}. This raises the question how we should {\em define\/} a universal distribution: what choices for $\ud{}{\gamma}$ are still `universal' (and define an MDL method) and what choices are not? Informally, every distribution $\ud{}{\gamma}$ that for no $z^n \in {\cal Z}^n$ has $\ud{}{\gamma}(z^n) \ll \ud{\nml}{\gamma}(z^n)$ is `universal' relative to $\cM_{\gamma}$. For parametric models such as exponential families, the `$\ll$' is partially formalized by requiring that at the very least, they should satisfy (\ref{eq:kover2}) below (G07 is much more precise on this).   
\item Third, we have not yet said how one should choose the `luckiness function' $v$ --- and one needs to make a choice to apply MDL in practice. The interpretation of $v$ is closely tied to the predictive interpretation of MDL, and hence we postpone this issue to the end of Section~\ref{sec:codelength}. 

\item Fourth, the motivation so far is incomplete --- we still need to explain why and how to incorporate the distribution $\pi$ on model index $\gamma$. This is done in Section~\ref{sec:large} below. 
\end{enumerate}

\subsection{Asymptotic Expansions}\label{sec:asymptotic}
Now let $\ud{}{}$ be defined relative to a single parametric model $\cM$. It turns out that all universal codes we mentioned have in common
that, for `sufficiently regular' $k$-dimensional parametric models, the log-likelihood for given data $z^n$
satisfies the following celebrated asymptotics, often called the {\em MDL\/} or {\em BIC\/} expansion: for all
`sufficiently regular' data sequences $z_1, z_2, \ldots$, there exists
a constant $C \in {\mathbb R}$ independent of $n$ such that for all $n$:
\begin{equation}\label{eq:kover2}
- \log \ud{}{}(z^n) \leq   
- \log p_{\mlest{\theta}(z^n)}(z^n) + 
\frac{k}{2} \log n +  C.
\end{equation}
For $\ud{\nml}{}$ and $\ud{\bayes}{}$, this holds for any choice of luckiness function $v$ and prior $w$ that is continuous and strictly positive on the parameter space $\Theta$. For $\ud{\twopart}{w}$, this holds for clever choices of the discretization $\ddot{\Theta}$ and the probability mass function $w$; for $\pud{\preq}{\breve\theta}{}$, this holds in a weaker expectation sense (see Section~\ref{sec:kotlowski}), as long as $\breve\theta$ is a suitably smoothed version of the ML estimator.  
Essentially, `sufficiently regular' parametric models are all exponential families (such as Bernoulli, multinomial, normal, gamma, beta, Poisson,...) and curved exponential families; corresponding results also hold for regression with (generalized) linear models. `Sufficiently regular data' are all sequences for which there is an INECCSI subset $\Theta_0$ of the parameter space $\Theta$ such that, for all large $n$, the ML estimator of the sequence lies within $\Theta_0$. Here INECCSI stands for a set whose  {\em I\/}nterior is {\em N}on-{\em E}mpty and whose {\em C\/}losure is a {\em C}ompact {\em S}ubset of the {\em I}nterior of $\Theta$. Essentially, this is any bounded subset of the same dimensionality as $\Theta$ that does not touch the boundaries of $\Theta$ itself; in the Bernoulli example, it would be any set of the form $[\epsilon, 1- \epsilon]$ for $\epsilon >0$. For all universal distributions considered except $\ud{\preq}{}$, as long as appropriate priors/estimators/luckiness functions are used,  (\ref{eq:kover2}) will hold uniformly for all sequences in any INECSSI subset $\Theta_0$, but the constant $C$ may grow larger if we replace $\Theta_0$ by a strictly larger INECCSI subset $\Theta'_0$ with $\Theta_0 \subsetneq \Theta'_0 \subsetneq \Theta$. (for $\ud{\preq}{}$ see Section~\ref{sec:kotlowski}). For the first four universal distributions, the inequality is actually equality up to a constant --- (\ref{eq:kover2}) also holds with $\leq$ replaced by $\geq$, for a different constant. For the switch distribution $\ud{\switch}{}$ however, the left-hand side will be significantly smaller
for a small but important subset of possible data sequences. 
Finally, since (\ref{eq:kover2}) thus also holds with $\ud{}{} = \ud{\nml}{}$ and $\leq$ replaced by $\geq$, 
exponentiating (\ref{eq:kover2}), we see that, if one restricts the
minimum in (\ref{eq:maximin}) to all such `sufficiently regular' $z^n$, $\text{\sc FR}(\ud{\bf u}{}(z^n))$ is guaranteed to be within a
constant (independent of $n$) factor of the optimal 
$\text{\sc FR}(\ud{\nml}{},z^n)$, for ${\bf u} \in \{ \bayes, \twopart,\preq, \switch,\ripr \}$. 

\paragraph{The NML/$\comp$ Expansion and the Jeffreys (Fisher Information) Integral} For the case that the model $\cM = \{p_{\theta} : \theta \in \Theta\}$ is a $k$-dimensional exponential family and $\ud{}{}$ is the NML or Bayes
distribution, we can be significantly more precise and evaluate the
constant $C$ in (\ref{eq:kover2}) up to $o(1)$: we get, under some weak additional regularity conditions on $\cM$ and $v$: 
\begin{align}
    \label{eq:nmlasymptotics}
\comp(\cM; v ) & \equiv  - \log \pud{\nml}{v}{}(z^n) - [- \log p_{\mdlest{\theta}{v}(z^n)} - \log v(\mdlest{\theta}{v}(z^n))] \nonumber \\ & = - \log \pud{\nml}{v}{}(z^n) - [- \log p_{\mlest{\theta}(z^n)} - \log v(\mlest{\theta}(z^n))] + o(1) \nonumber \\ & =   
 \frac{k}{2}\log {\frac{n}{2\pi}} + \int_\Theta v(\theta) \cdot \sqrt{|I(\theta)|} d\theta + o(1),
\end{align}
where $k$ is the dimension of the model, $|I(\theta)|$ is the determinant of the $k \times k $ Fisher information matrix at parameter $\theta$, the integral is over the parameter space $\Theta$, and the remainder term $o(1)$ vanishes as the sample size grows unbounded.
This was first shown (essentially) by \cite{Rissanen96},
for the case that  $\Theta$ is restricted to an INECCSI subset of the full parameter space (so that $\ud{\nml}{}$ with $v \equiv 1$ is defined), and $v \equiv 1$. For this uniform $v$ case,
\cite{MyungBP00} gave a differential geometric interpretation of the Fisher information term, relating it to an intrinsic `volume' of the parameter space. The general result for nonuniform $v$, and without INECCSI restrictions, was very recently shown in a breakthrough paper by \cite{suzuki2018exact}, solving Open Problem 6 from G07.

Analogously  to $\ud{\nml}{}$ (in fact much easier mathematically), we can expand $\ud{\bayes}{}$ using a classical Laplace approximation; under the same conditions as before, with now the additional restriction that there exists an arbitrary INECCSI subset $\Theta_0$ of $\Theta$ such that for all large $n$, the  data have ML estimator within $\Theta_0$, we find that
\begin{equation}\label{eq:bayesasymptotics}
 - \log \pud{\bayes}{w}{}(z^n) =  -\log p_{\mlest{\theta}(z^n)}(z^n)  + \frac{k}{2}\log \frac{n}{2\pi} + 
\frac{1}{2}  \log |I(\mlest{\theta}(z^n))| - \log w(\mlest{\theta}(z^n)) + o(1).
\end{equation}
From (\ref{eq:nmlasymptotics}) and (\ref{eq:bayesasymptotics}) we see that, if the {\em generalized Jeffreys integral\/} $\int v(\theta) \cdot \sqrt{|I(\theta)|} d \theta$ is finite (see Appendix~\ref{app:nml}), then 
there is a special choice of prior $w$, the {\em generalized Jeffreys prior\/}, with $w(\theta) = v(\theta)  \sqrt{| I(\theta)|} / \int v(\theta)  \sqrt{ |I(\theta)|} d \theta$, under which $- \log \ud{\nml}{}(z^n)$ and $- \log \ud{\bayes}{}(z^n)$ do not just coincide up to $O(1)$, but become asymptotically indistinguishable. If $v \equiv 1$, this $w$ coincides with the well-known {\em Jeffreys' prior\/} $\jp$ popular in Bayesian inference; the special case of this prior for the Bernoulli model was encountered in Example~\ref{ex:bernoulli}. Thus, the Bayesian universal distribution with the (generalized) Jeffreys' prior can be a very good alternative of $\pud{\nml}{v}{}$.

\subsection{Unifying Model Selection and Estimation}
\label{sec:large}
Suppose we are given a countable collection of models $\{\cM_{\gamma} : \gamma \in \Gamma\}$.
%\bigcup_{\gamma \in \Gamma} \cM_{\gamma}.  
Recall that the basic idea above was to associate each individual model $\cM_{\gamma}$ with a single distribution $\ud{}{\gamma}$. It seems reasonable to do the same at the level of `meta-parameters' $\gamma$: we set $\overline{\cM } := \{ \ud{}{\gamma}: \gamma \in \Gamma\}$
%
%with the union of all models under consideration, which, of course, may also %be viewed as a single huge model. Thus, 
and in complete analogy to (\ref{eq:nmluniversal}), we define {\em the meta- universal distribution}
\begin{equation}\label{eq:large}
\pud{\nml}{\pi}{}(z^n) := \frac{\max_{\gamma \in \Gamma} \ud{}{\gamma}(z^n ) \pi(\gamma)}{\int_{z^n} \max_{\gamma \in \Gamma} \ud{}{\gamma}(z^n) \pi(\gamma) d z^n}
\end{equation}
for some nonnegative weight function $\pi$ on $\Gamma$.
It then makes sense to postulate that the best sub-model $\cM_{\gamma}$ for the given data $z^n$ is given by the $\gamma$ achieving the maximum in (\ref{eq:large}). Note that for determining this maximum, the denominator in (\ref{eq:large}) plays no role. 

Let us assume that all $\ud{}{\gamma}$ have already been defined. Then we can use any $\pi$ such that the overarching $\pud{\nml}{\pi}{}$ in (\ref{eq:large}) exists. We can now formulate a {\em general MDL Principle for model selection\/}: we start with a (potentially huge) set of candidate distributions $\cM_{\full}$. We next carve up $\cM_{\full}$ into interesting {\em sub-models\/} $\cM_{\gamma}$ with $\gamma \in \Gamma$, so that $\bigcup_{\gamma \in \Gamma} \cM_{\gamma} = \cM_{\full}$. We then associate each $\cM_{\gamma}$ with a universal distribution $\ud{}{\gamma}$, and we equip $\overline{\cM}$ as defined above with luckiness function $\pi$ (note that $\cM_{\full}$, a countable union of (usually) uncountable sets, consists of all distributions under consideration, while $\overline{\cM}$ is a countable set). We then base the selection of a sub-model $\cM_{\gamma}$ on  (\ref{eq:large}). What we earlier called  the `general MDL Principle' underneath (\ref{eq:selectingk}) was the special case in which $\sum \pi(\gamma) = 1$, i.e. $\pi$ is a probability mass function. Via (\ref{eq:twopartone}) we see that for any such probability mass function $\pi$, the denominator in (\ref{eq:large}) is well-defined, hence $\pi$ is a valid luckiness function. 

Now consider the special case in which 
every $\ud{}{\gamma}$ is chosen to be an  NML distribution $\pud{\nml}{v}{\gamma}$ for some luckiness functions $v_{\gamma}$.
We take some function $\pi': \Gamma  \rightarrow \reals^+_0$ (which we will relate to the $\pi$ above later on) and we set, for $\theta \in \Theta_{\gamma}$,  $v_{\full}(\gamma,\theta) := \pi'(\gamma) v_{\gamma}(\theta)$.  
We can use $\pud{\nml}{v_{\full}}{}$ for parameter estimation on the joint parameters $(\gamma,\theta)$ just as we did earlier for parametric models, by using the MDL estimator $\widehat{(\gamma,\theta_{\gamma})}_{v_{\full}}$ picking the $(\gamma,\theta_{\gamma})$ minimizing, over $\gamma \in \Gamma, \theta_{\gamma} \in \Theta_{\gamma}$,  
\begin{multline}\label{eq:combimdl}
- \log p_{\theta_{\gamma}}(z^n) - \log v_{\full}(\gamma,\theta_{\gamma}) + \comp(\cM_{\full},v_{\full})
= \\
- \log p_{\theta_{\gamma}}(z^n) - \log v_{\gamma}(\theta) - \log \pi'(\gamma) + \comp(\cM_{\full},v_{\full})
\end{multline}
where again $\comp(\cM_{\full},v_{\full})$ plays no role in the minimization. This MDL estimator really combines model selection (estimation of $\gamma$) and parametric estimation (estimation of $\theta_\gamma$). If we now define $\pi(\gamma) := \pi'(\gamma)/\exp(\comp(\cM_{\gamma},v_{\gamma}))$, we find that $\pud{\nml}{\pi}{}$ defined relative to model $\overline{\cM}$ as in (\ref{eq:large}) is equal to $\pud{\nml}{v_{\full}}{}$ defined relative to the full union of models $\cM_{\full}$, and the $\gamma$ achieving the maximum in (\ref{eq:large}) coincides with the $\gamma$ minimizing (\ref{eq:combimdl}). 
This indicates that model selection and estimation is really the same thing with MDL: if we are given a single parametric model $\cM_{\gamma}$ with luckiness $v_{\gamma}$, we pick the $\theta$ minimizing the first two terms in (\ref{eq:combimdl}) for fixed $\gamma$; if we are interested in both $\theta$ and $\gamma$, we minimize over all terms; and if we are only interested in $\gamma$, we pick the $\gamma$ achieving the maximum in (\ref{eq:large}), which, by construction, will give us the same $\gamma$ as the joint minimization over (\ref{eq:combimdl}). 
\paragraph{Two-Part vs. One-Part Codes: the Role of Data Compression}
In the oldest [1978] version of MDL, only two-part codes on countable sets were used: the minimum over $\theta \in \Theta_{\gamma}$ was taken over a discretized grid $\ddot{\Theta}_{\gamma}$ and $v_{\gamma}$ was a probability mass function over this grid; then for all $\gamma$,  $\comp(\cM_{\gamma},v_{\gamma}) \leq 0$ and $\comp(\cM,v) \leq 0$ (see (\ref{eq:twopartone})) and they were both approximated by $0$. 
From the {\em Kraft inequality\/} \citep{CoverT91} we see where the name `2-part code' comes from: this inequality says that for every probability mass function $\pi$ on a countable set ${\cal A}$, there exists a lossless code such that for all $a \in {\cal A}$, the number of bits needed to encode $a$, is given by $- \log \pi(a)$. Thus the resulting method can be interpreted as picking the $(\ddot{\theta},\ddot{\gamma})$ minimizing the two-stage code length of the data, where first the parameters $(\theta,\gamma)$ are encoded using $- \log \pi({\gamma}) - \log v_{\gamma}(\theta)$ bits, and then $z^n$ is encoded `with the help of $\gamma$', using $- \log p_{\theta_{\gamma}}(z^n)$ bits (in fact, the encoding of $(\gamma,\theta)$ itself has two sub-stages here so we really have a two-part code where the first part itself has two parts as well). 

The discretization involved in using  a probability mass function/code for continuous-valued $\theta$ makes things (unnecessarily, as was gradually discovered over the last 30 years) very complicated in general. Also, if one combines choice of $\theta$ with choice of $\gamma$, the approximation of  $\comp(\cM_{\gamma},v_{\gamma})$ as $0$ introduces some suboptimalities. Thus, one would like to code the data in a way that avoids these two issues. It turns out that this can be achieved by replacing 2-part by 1-part codes for the data, namely, to use codes with length $- \log \pud{\nml}{v}{}(z^n)$: assuming for simplicity that data are discrete, the same Kraft inequality implies that there must also be a code, directly defined on $z^n$, which achieves  code length for each $z^n$ given by  $- \log \ud{\nml}{}(z^n)$. Thus, even though for general luckiness functions $v$ this code length cannot be decomposed into two sub-code lengths, it remains a valid code length and the name {\em MDL\/} for the resulting procedure remains, we feel, justified. In the past, it was sometimes thought by some MDL fans that 2-part codes on countable sets would somehow lead to inherently better estimates $\mdlest{\theta}{w}$ than estimators $\mdlest{\theta}{v}$ for general luckiness functions as in (\ref{eq:brevepre}). However, after 30 years it turned out  there is nothing either conceptually or mathematically that indicates the need for 2-part codes and countable sets: for any luckiness function $v$, the resulting procedure has a code length interpretation, and  (\cite{GrunwaldM19} show that all consistency and convergence results that hold for 2-part estimators also hold for general MDL estimators (Section~\ref{sec:mehta}) --- thus invalidating the conjecture in Open Problem 13 of G07 that postulated a special status for luckiness functions $v$ that are probability mass functions on countable sets.  
For the luckiness function $\pi$ on the discrete structure $\Gamma$ however, it {\em is\/} quite reasonable to choose a probability mass function: no probability mass is wasted (since the denominator in (\ref{eq:large}) plays no role in choosing $\gamma$), and designing $\pi$ by thinking about the code lengths $- \log \pi(\gamma)$ comes very naturally, as the following example illustrates:  . 
\begin{example}{\bf [Variable Selection - $L_1$ vs $L_0$-penalties]} \label{ex:varsel} Suppose that each data point $Z_i = (X_i,Y_i)$ where $Y_i$ denotes the variable to be predicted and  $X_i = (X_{i1},\ldots,X_{im}) \in \reals^m$ is a vector of covariates or `features' that may or may not help for predicting $Y_i$. We consider a linear model $\cM = \{ p_{\beta}: \beta \in \reals^m\}$, decomposed into sub-models $\cM_{\gamma}$ expressing 
$$
Y_i = \sum_{j=1}^m \gamma_j \beta_j X_{ij} + \epsilon_i,
$$
where $\epsilon_1, \epsilon_2, \ldots$ represents $0$-mean i.i.d. $N(0,\sigma^2)$ normally distributed noise and  $\gamma = (\gamma_1, \ldots, \gamma_m) \in \{0,1\}^m$ is a binary vector indicating which variables are helpful for predicting the $Z_i$. Thus, if $\gamma$ has $k$ 0-components, then $\cM_{\gamma}$ is effectively an $m-k$ dimensional model. Our task is to learn, from the data, the vector $\gamma^*$ indicating which variables are truly relevant, and/or such that predictions of new $Y$ given new $X$ based  on $\cM_{\gamma^*}$ are as good as possible. 

In light of the above, a straightforward way to use MDL here is to pick the $\gamma$ minimizing 
\begin{equation}\label{eq:regmdl}
- \log \pud{\nml}{v}{{\gamma}}(y^n\mid x^n) - \log \pi(\gamma),\end{equation}
where we refer to G07, Figure 14.2 for an explanation why we can condition on $x^n$ here. $v_{\gamma}$ in (\ref{eq:regmdl}) is an appropriately chosen luckiness function and  $\pi$  is really a probability mass function, such that $L_{\pi}(\gamma) := - \log \pi(\gamma)$ can be interpreted as the number of bits needed to encode $\gamma$ using a particular code. In terms of coding, a natural choice for such a code would be to first encode the number of nonzero components $k_{\gamma}$ in $\gamma$ using a uniform code (that assigns equal code length to all possibilities). Since $0 \leq k_{\gamma} \leq m$, there are $m+1$ possibilities, so this takes $\log_2 (m+1)$ bits. In a second stage, one encodes the location of these components. There are $\binom{m}{k_{\gamma}}$ possibilities here, so this takes $\log_2 \binom{m}{k_{\gamma}}$ bits using a uniform code. All in all, one needs 
\begin{equation}\label{eq:secondpart}
\log (m+1) + \log  \binom{m}{k_{\gamma}}
\end{equation}
`nits' (bits re-expressed in terms of natural logarithm $\log$) to encode $\gamma$. Then (\ref{eq:secondpart}) can be written as $- \log \pi(\gamma)$, with  $\pi(\gamma) = 1 / ((m +1)\cdot \binom{m}{k_{\gamma}})$, which, as predicted by the Kraft inequality, sums to $1$ over $\gamma \in \{0,1\}^m$.  

As to the left part of the code length (\ref{eq:regmdl}), if the variance $\sigma^2$ is known, a natural luckiness function $v_{\gamma}$ to use is a $m- k_{\gamma}$-dimensional Gaussian with mean $0$ and variance $\sigma^2 \Sigma$ for some (usually diagonal) covariance matrix $\Sigma_{\gamma}$. This gives (see Chapter 12 of G07)
\begin{multline}\label{eq:firstpart}
    - \log \pud{\nml}{v}{{\gamma}}(y^n \mid x^n) = \\ \frac{1}{2 \sigma^2} \sum_{i=1}^n \left(y_i - \sum_{j=1}^m \mlestgamma{\beta}{j} \gamma_j x_i \right)^2 + \frac{n}{2} \log {2}{\pi} \sigma^2 + \frac{1}{2} \log \left| {\bf X}^T {\bf X} + \Sigma^{-1} \right|
    + \frac{1}{2} \log \left| \Sigma \right|, \end{multline}
where ${\bf X} = (X_1, \ldots, X_n)$ and $| \cdot |$ stands for determinant. We thus end up finding the $\gamma$ minimizing the sum of (\ref{eq:firstpart}) and (\ref{eq:secondpart}). If, as here, the noise is normal with fixed variance, then for this choice of luckiness function, $\pud{\nml}{v}{{\gamma}}(y^n \mid x^n)$ actually {\em coincides\/} with $\pud{\bayes}{w}{{\gamma}}(y^n \mid x^n)$ for a particular prior $w_{\gamma}$, thus one has a Bayesian interpretation as well (\cite{bartlett2013horizon} show that such a precise correspondence between NML and Bayes only holds in quite special cases, see Appendix~\ref{app:nml}). If the variance $\sigma^2$ is unknown, one can treat it as a nuisance parameter and equip it with the improper Haar prior, leading to a modification of the formula above; see Example~\ref{ex:haar}. 
Even if the noise is not known to be normally distributed, one can often still use the above method --- pretending the noise to be normally distributed and accepting that one is misspecified --- by varying the {\em learning rate}, as briefly explored in Section~\ref{sec:misspecification}. 

Note that the code/prior we used here induces {\em sparsity\/}: if there exists a $\gamma$ with mostly 0 components that already fits the data quite well, we will tend to select it, since, for  $k \ll n$, $\log \binom{n}{k}$ increases approximately linearly in $k$. {\em  That does not mean that we necessarily believe that the `truth' is sparse --- it just expresses that we hope that we can already make reasonably good predictions with a small number of features}.  

An alternative, and very popular, approach to this problem is the celebrated {\em Lasso\/} \citep{HastieTF01}, in which we consider only the full model $\cM = \cM_{(1,1,\ldots, 1)}$ and we pick the $\beta \in \reals^m$ minimizing
\begin{equation}\label{eq:lasso}
\frac{1}{2 \sigma^2} \sum_{i=1}^n \left(y_i - \sum_{j=1}^m \beta_j \gamma_j x_i \right)^2 + \frac{\lambda}{2 \sigma^2} \sum_{j=1}^m |\beta_j|
\end{equation}
for some regularization parameter $\lambda > 0$ (the factor $\sigma^2$ plays no role in the minimization; it is incorporated only to facilitate comparison with (\ref{eq:firstpart}). It is known this will tend to select $\beta$ with many zero-components, thus also inducing sparsity, and it can be implemented computationally much more efficiently than the two-part approach sketched above, effectively replacing $L_0$ by $L_1$-penalties.
With our `modern' view of MDL, this can be thought of as a form of MDL too, where we simply impose the luckiness function 
$v(\beta)= \exp(- (\lambda/\sigma^2) \sum_{j=1}^m | \beta_j |)$ and use the estimator $\mdlest{\theta}{v}$ given by (\ref{eq:brevepre}). The luckiness function $v$ depends on $\lambda$; the optimal choice of $\lambda$ is then once again related to the optimal learning rate; see Section~\ref{sec:misspecification}. Finally, we note that there exists a third MDL approach one can use here: one starts out with an NML approach similar to (\ref{eq:regmdl}) but then performs a continuous relaxation of the resulting optimization problem; the resulting `relaxed' NML criterion is then once again tractable and similar to an $L_1$-optimization problem such as (\ref{eq:lasso}); this approach has been described by \cite{MiyaguchiY18} who extend the idea to group lasso and other settings. 
\end{example}
\subsection{Log-Loss Prediction and Universal Distributions}
\label{sec:codelength}
Now consider the simple case again with a finite set of models $\{ \cM_{\gamma}: \gamma \in \Gamma \}$ where $\Gamma$ is small compared to $n$ and we use the uniform prior $\pi$, picking the $\gamma$ maximizing $\ud{}{\gamma}$. It was the fundamental insight of \cite{Rissanen84} and \cite{Dawid84}
that such model choice by maximizing  $\ud{}{\gamma}(z^n)$ for a
single distribution $\ud{}{\gamma}$ can be motivated in a different
way as well --- in essence, it selects the model with the best predictive
performance on unseen data. This approach shows that MDL is quite
similar in spirit to cross-validation, the main difference with
leave-one-out cross validation being that the {\em cross\/} in
cross-validation is replaced by a {\em forward\/} and that the loss
function used to measure prediction error is restricted to be the
logarithmic score, also commonly known as {\em log loss\/} (which, however, is often used in cross-validation
as well).

Formally, the log loss of predicting a single outcome $z \in
\cZ$ with a distribution $p$ is defined as $- \log p(z)$: the larger
the probability density, the smaller the loss. If one predicts a
sequence of $n$ outcomes $z^n = (z_1, \ldots, z_n)$ with $n$
predictions $p_1, p_2, \ldots, p_n$, then the {\em cumulative
  log loss\/} is defined as the sum of the individual losses:
$\sum_{i=1}^n - \log p_i(z_i)$.

Now, if we adopt a probabilistic world view and represent our beliefs about $z^n$ by a probability distribution $\ud{}{}$, then the obvious way to make sequential predictions is to set $p_i := \ud{}{}(Z_i = \cdot \mid z^{i-1})$, so that $-\log p_i(z_i) = - \log \ud{}{}(z_i \mid z^{i-1})$. For arbitrary probability distributions, we have, by the formula for conditional probability: for all $z^n \in \cZ^n$,  $p(z^n) = \prod_{i=1}^n p(z_i \mid z^{i-1})$. Taking logarithms gives: 
\begin{equation}\label{eq:phil}
\sum_{i=1}^n - \log \ud{}{}(z_i \mid z^{i-1}) = - \log \ud{}{}(z^n).
\end{equation}
In words, for every possible sequence, {\em the cumulative log
  loss obtained by sequentially predicting $z_i$ based on the
  previously observed data $z^{i-1}$ is equal to the minus
  log-likelihood}.

Conversely, if we are given an arbitrary {\em sequential prediction
  strategy\/} $\bar{s}$ which, when input a sequence $z^{i-1}$ of
arbitrary length $i-1$ outputs a prediction for the next outcome $z_i$
in the form of a probability distribution $\bar{s}_{z^{i-1}}$ on
$\cZ$, we can {\em define\/} $\ud{}{}(z_i \mid z^{i-1}) :=
\bar{s}_{z^{i-1}}(z_i)$ and then further define $\ud{}{}(z^n) :=
\prod_{i=1}^n \ud{}{}(z_i \mid z^{i-1})$. A simple calculation shows
that we must have $\int_{z^n \in \cZ^n} \ud{}{}(z^n) d z^n  = 1$, so we have {\em constructed\/} a probability distribution
$\ud{}{}$ which once again satisfies (\ref{eq:phil}).  The fundamental
insight here is that, when the log loss is used, {\em every
  probability distribution defines a sequential prediction strategy
  and --- perhaps more surprisingly --- vice versa, every sequential
  prediction strategy defines a probability distribution, such that on
  all sequences of outcomes, the minus log likelihood is equal to the
  cumulative loss}.
\begin{example} 
Consider again the Bernoulli model $\cM= \{ p_{\theta}: \theta \in [0,1] \}$. Each element $p_{\theta} \in \cM$ defines a prediction strategy which, no matter what happened in the past, predicts that the probability that the next outcome $Z_i = 1$ is equal to $\theta$. It incurs cumulative loss, on sequence $z^n$ with $n_1$ ones and $n_0 = n- n_1$ zeros, given by $n_1 (- \log \theta) + n_0 (- \log (1- \theta))$. The Bayesian universal distribution $\pud{\bayes}{\jp}{}$ with Jeffreys' prior that we considered in Example~\ref{ex:bernoulli} satisfies, as was already mentioned, $\pud{\bayes}{\jp}{}(Z_{m+1} = 1 \mid z^{m}) = (m_1 + (1/2))/(m+1)$, so it `learns' the probability of $1$ based on past data and does not treat the data as i.i.d. any more. The asymptotic expansion (\ref{eq:bayesasymptotics}) then shows that its cumulative loss is of order 
\begin{equation*}
- \log p_{\mlest{\theta}(z^n)}(z^n) + \frac{1}{2} \log n + O(1) = - n_1 \log (n_1/n) - n_0 \log (n_0/n)  + \frac{1}{2} \log n + O(1). 
\end{equation*}
\end{example}
We may now ask: given a parametric model ${\cal M}$, what distribution
(i.e. prediction strategy) in ${\cal M}$ leads to the best predictions
of data $z_1, \ldots, z_n$? For simplicity we will assume that data
are i.i.d. according to all $p_{\theta} \in {\cal M}$.  We have to
distinguish between the best sequential prediction strategy {\em with
  hindsight} and the best prediction strategy that can be formulated
before actually seeing the data. The former is given by the
$p_{\theta} \in {\cal M}$ achieving
$$
\min_{\theta \in \Theta} 
\sum_{i=1}^n - \log p_{\theta}(z_i \mid z^{i-1})  = \min_{\theta \in \Theta} - \log p_{\theta}(z^n) = - \log p_{\mlest{\theta}(z^n)}(z^n),
$$
i.e. the best predictions with hindsight are given by the ML
distribution $\mlest{\theta}(z^n)$. However, $\mlest{\theta}(z^n)$ is only knowable after seeing all the data $z^n$, whereas in reality, we have, at each time $i$, to make a prediction $\ud{}{}(Z_i \mid z^{i-1})$ relying only on the previously seen data $z^{i-1}$.  We might thus aim for a prediction strategy (distribution) $\ud{}{}$ which will tend to have a small {\em regret\/} (additional prediction error) 
\begin{multline}
    \label{eq:simon}
\regret(\ud{}{},z^n) = \sum_{i=1}^n - \log \ud{}{}(z_i \mid z^{i-1}) - \left[
\min_{\theta \in \Theta} \sum_{i=1}^n - \log p_{\theta}(z_i \mid z^{i-1}) \right] \\
= -\log \ud{}{}(z^n) + \log p_{\mlest{\theta}(z^n)}(z^n).
\end{multline}
But what does `tend' mean here? One strict way to implement the idea is to require (\ref{eq:simon}) to be small in the worst case --- one looks for the distribution $\ud{}{}$ achieving
\begin{equation}\label{eq:garfunkel}
\min_{\ud{}{}} \max_{z^n \in \cZ^n} \regret(\ud{}{},z^n),
\end{equation}
where the minimum is over all probability distributions on $\cZ^n$.  But comparing (\ref{eq:simon}) and (\ref{eq:garfunkel}) with
(\ref{eq:maximin}), using that $- \log$ is strictly decreasing, we see that the $\ud{}{}$ achieving (\ref{eq:garfunkel}) is just the NML distribution with $v \equiv 1$, which was already our `favourite' distribution to use in MDL model comparison any way! And, just like before, if (\ref{eq:garfunkel}) has no solution, we may add a $- \log v$ luckiness term to (\ref{eq:simon}) so as to regularize the problem, and then the optimal prediction strategy will be given by $\pud{\nml}{v}{}$. We also see that with $v \equiv 1$, $\comp(\cM,v)$ is equal to the minimax regret (\ref{eq:garfunkel}); and with nonuniform $v$, $\comp(\cM)$ will become equal to the minimax {\em luckiness regret}, i.e. (\ref{eq:simon}) with a $- \log v$ term added. 

We now also see where the idea to use the prequential plug-in distribution $\ud{\preq}{}$ instead of $\ud{\nml}{}$ comes from: if calculating $\pud{\nml}{v}{}$ is too difficult, or if the {\em horizon\/} $n$ (which is needed to calculate $\pud{\nml}{v}{}$) is unknown, we might simply pick any estimator $\breve\theta$ which we think is `reasonable' and replace our prediction $\pud{\nml}{v}{}(Z_i \mid z^{i-1})$ by $p_{\breve\theta(z^{i-1})}(Z_i)$ --- if the estimator was chosen cleverly, we can expect the resulting cumulative regret to be small. Reconsidering (\ref{eq:kover2}), we see that all the universal distributions, viewed as prediction strategies, with the right choice of luckiness functions, priors and/or estimates, can be made to achieve a logarithmic (in $n$) worst-case regret --- since the cumulative log-loss achieved by the best predictor in hindsight usually grows linearly in $n$, a logarithmic regret is quite satisfactory. Returning to our Bernoulli example, we see that the cumulative log loss obtained by $\mlest{\theta}$, the best with  hindsight, is equal to $n H(n_1/n)= n H(\mlest{\theta}(z^n))$, where $H(\theta)$ is the binary entropy, $H(\theta) := - \theta \log \theta - (1-\theta) \log (1- \theta)$. Note that, in line with the above discussion,  $n H(\mlest{\theta})$  is linear in $n$ unless $\mlest{\theta}$ tends to $0$ or $1$, but the regret of $\ud{\bayes}{}$ with Jeffreys' prior is logarithmic in $n$. 

We thus get a novel interpretation of MDL: it associates each  model $\cM_{\gamma}$ with a sequential prediction strategy $\ud{}{\gamma}$ that is  designed to achieve small regret compared to the hindsight-optimal prediction strategy within $\cM_{\gamma}$; it then picks the model for which the corresponding prediction strategy achieves the smallest cumulative loss on the data. 

\paragraph{Related Work} Dawid (see \citep{Dawid84} and many
subsequent works) suggests to use this {\em prequential model choice\/}
also with respect to loss functions other than the logarithmic loss;
minimax optimal cumulative prediction strategies without making
stochastic assumptions about the data, with log loss but (mainly) with
other loss functions, are one of the main topics in {\em machine
  learning theory\/}; see for example \cite{CesaBianchiL06}; but there
they are generally not used for model comparison or selection.

\paragraph{Why the logarithmic score?}
Why does it make sense to minimize cumulative log loss? Outside of the MDL world, the log loss is often used for two reasons: first, it is (essentially) the only {\em local proper scoring rule\/} \citep{dawid2007geometry}. Second, it has an interpretation in terms of {\em money\/}: for every sequential prediction strategy, there is a corresponding `sequential investment' strategy such that, the smaller the cumulative log-loss, the larger the monetary gains made with this strategy (`Kelly Gambling', \citep{CoverT91,GrunwaldHK19}). 

Within the MDL field however, the use of the log loss comes from the Kraft inequality, which directly relates it to {\em lossless data compression}. As we already saw before Example~\ref{ex:varsel}, for any sequential prediction strategy, i.e. every distribution $p$ on sequences of length $n$, there is a {\em lossless code\/} $C$ such that, for all sequences of length $n$, 
$$ - \log_2 p(z^n) = \text{\sc nr. of bits needed to code the data $z^n$  using code $C$}. $$
Conversely, for any code $C$, there is a corresponding distribution $p$ such that the above holds (see Chapter 3 of G07 for a very extensive explanation). Thus, the original MDL idea to `take the model that compresses the data most' is first made more formal by replacing it by `associate each model with a code that compresses well whenever some distribution in this model compresses well', and this turns out to be equivalent to `associate each model $\cM_{\gamma}$ with a distribution $\ud{}{\gamma}$  that assigns high likelihood whenever some distribution in the model assigns high likelihood'. 

\paragraph{MDL Prediction and `Improper' Estimation}
As is clear from the prequential interpretation of MDL given above, once a universal distribution $\ud{}{}$ has been fixed, one can use it to {\em predict\/} $Z_i$ given $z^{i-1}$ by $\ud{}{}(Z_i \mid z^{i-1})$. At least  for i.i.d. data, we can estimate the underlying `true' distribution $p^*$ based on such predictions directly, by simply interpreting $\ud{}{}(Z_i \mid z^{i-1})$ as an estimate of $p^*$! 
This is different from the previous form of MDL estimation described in Section~\ref{sec:large}, which was based on MDL (penalized ML) estimators $\mdlest{\theta}{v}$. 
Note that this standard MDL estimator is `in-model' or {\em proper\/} (to use machine learning terminology \citep{shalev2014understanding}), whereas $\ud{}{}(Z_i \mid z^{i-1})$ is {\em out-model\/} or {\em improper\/}: in general, there may be no $p \in \cM$ such that $\ud{}{}(\cdot \mid z^{i-1}) = p$. For example, with Bayes universal distributions, $\pud{\bayes}{w}{Z_i \mid z^{i-1}}$  will be a  mixture of distributions in $\cM$ rather than a single element; see G07 for more discussion. 
\subsection{The Luckiness Function}\label{sec:luckiness}
The choice of a luckiness function is somewhat akin to the choice of a prior in Bayesian statistics, yet --- as explained at length in Chapter 17 of G07 --- there are very important differences, both technically (luckiness functions are not always integrable) and philosophically.   Basically, a luckiness function just determines for what type of data one will be ``lucky'' ($v(\mdlest{\theta}{v}(z^n))$ large) and get small cumulative regret based on small samples (and presumably, good model selection results as well), and for what data one will be less lucky and get good results only when the data set grows much larger --- $v$ may thus be chosen for purely pragmatic reasons. For example, as in Example~\ref{ex:varsel} (see the italicized text there), if one assigns a large value $\pi(\gamma)$ to some model $\cM_{\gamma}$ within  a large collection of models $\{ \cM_{\gamma}: \gamma \in \Gamma\}$ where $\gamma$ is {\em sparse}, one may do this because one hopes that this sub-model will already lead to {\em reasonable\/} predictions of future data, even though one feels that, at the same time, when more data becomes available, a model $\cM_{\gamma'}$  with a much larger number of nonzero parameters may at some point almost certainly become better (Example~\ref{ex:varsel}). Such an interpretation is not possible with a Bayesian prior $\pi$, where a large value of $\pi(\gamma)$ indicates a strong belief that $\cM_{\gamma}$ is true, or at least, that predictions based on acting as if it will be true will be optimal --- a Bayesian with high prior on $\gamma$ considers $\cM_{\gamma}$ {\em likely\/} rather than just {\em useful} -- a distinction worked out in detail by \cite{grunwald2018safe}. Nevertheless, just like a Bayesian prior, the luckiness function has to be chosen by the user/statistician, and often contains a subjective element. Still, in contrast to Bayesian priors, since we invariably take a worst-case log-loss stance in MDL, there often is a uniquely preferable choice of luckiness function $v$ for parametric models $\cM$. First, if $\comp(\cM,v)< \infty$ with uniform $v$ and no clear prior knowledge or preference is available, then uniform $v$ is usually preferable over other $v$, since it achieves the worst-case optimal prediction performance. Second, if $\comp(\cM_{\gamma},v)= \infty$ with uniform $v$ for some $\gamma \in \Gamma$ we can often still set the first few, say $m$, outcomes aside and pick a luckiness function $v_{\gamma}(\theta) :=  p_{\theta}(z_1, \ldots, z_m)$ for $\theta \in \Theta_{\gamma}$. The corresponding estimator (for fixed $\gamma$) $\mdlest{\theta}{v_{\gamma}}$ based on data $z_{m+1}, \ldots, z_n$ as given by  (\ref{eq:brevepre}) will then be equal to the ML estimator based on the full data, $\mdlest{\theta}{v_{\gamma}}(z_{m+1}^n) = \mlestgamma{\theta}{\gamma}(z^n)$, and by choosing $m$ large enough, one can often get that $\comp(\cM_{\gamma},v_{\gamma})$ is finite after all for all $\gamma \in \Gamma$ and one may then compare models by picking the $\gamma$ maximizing $\pud{\nml}{v}{\gamma}(z_{m+1}, \ldots, z_n)$. Thus, the luckiness function is now determined by the first few `start-up' data points, and one uses NML based on the remaining data points with an estimator that coincides with the ML estimator based on the full data. G07 argues why  this data-driven luckiness function is the best default choice available; note that it is analogous to our use of improper Bayes priors as described in Example~\ref{ex:gauss}. 
\section{Novel Universal Distributions}
\subsection{The Switch Distribution and the AIC-BIC Dilemma}
\label{sec:switch}
The {\em AIC-BIC dilemma\/} (see for example \citep{Yang05a} and the
many citations therein) is a classic conundrum in the area of
statistical model selection: if one compares a finite number of
models, the two standard benchmark methods, with (different) asymptotic
justifications, are AIC and BIC. Suppose one first selects a model
using AIC or BIC. One then predicts a
future data point based on, e.g., maximum likelihood estimation, or
by adopting the Bayes/MDL predictive distribution, within the chosen
model. If one compares a finite number of models, then AIC tends to
select the one which is optimal for prediction (compared to BIC, the
predictions converge faster, by a factor of order $\log n$, to the
optimal ones). On the other hand, BIC is consistent: with probability
1, it selects the smallest model containing the true distribution, for
all large $n$; the probability that AIC selects an overly large model
does not go to $0$ for large $n$. Both the predictive optimality and the consistency propertiy are desirable, but, like AIC and BIC, 
common methods all fail on one of the two. For example, MDL, with each of the four classical
distributions, and Bayes factor model selection will behave like BIC for large $n$ and be consistent but prediction-suboptimal; for any fixed $k$, leave-$k$-out and $k$-fold cross-validation
will tend to behave like AIC and have the reverse
behaviour. \cite{Yang05a} shows that, in general, this dilemma cannot
be solved: every consistent method has to be slightly prediction
suboptimal in some situations; he also shows that prediction by model averaging cannot solve this dilemma either. 

Nevertheless, as first shown by \cite{ErvenGR07} (who thereby solved Open Problem 8 and 17 of G17), one can design universal distributions
that `almost' get the best of both worlds: basing MDL model selection on them using (\ref{eq:selectingk}) one gets a criterion which is strongly
consistent while at the same time losing only an exceedingly small
order $\log \log n$ factor in terms of prediction quality compared to the
AIC-type methods. Although it can be applied to arbitrarily large
model collections, the idea of this so-called {\em switch
  distribution\/} $\ud{\switch}{}$ is best explained by
considering the simplest case with just two nested models $\cM_0
\subset \cM_1$: one starts with two standard universal distributions
(say, Bayesian or luckiness-NML) $\ud{}{0}$ for $\cM_0$ and $\ud{}{1}$ for $\cM_1$. For every $i > 0$, $\ud{}{1}$ defines a conditional distribution $\ud{}{1}(z_i, \ldots, z_n \mid z^{i-1})$. One now picks a `prior' distribution $\pi$ on the integers (typically one that decreases polynomially, e.g. $\pi(i) = 1/(i(i+1))$), and one defines a new universal distribution for $\cM_1$ by:
$$
\ud{\switch}{}(z^n) := \sum_{i=1}^n \pi(i) \ud{}{1}(z_i, \ldots, z_n \mid z^{i-1}) \cdot \ud{}{0}(z^{i-1}).
$$
This distribution is best understood from the prequential interpretation of MDL (Section~\ref{sec:codelength}). It will satisfy
\begin{align*}
\sum_{i=1}^n - \log \ud{\switch}{}(z_i \mid z^{i-1})
& = - \log \ud{\switch}{}(z^n) = - \log \sum_{i=1}^n \pi(i) \ud{}{1}(z_i, \ldots, z_n \mid z^{i-1}) \ud{}{0}(z^{i-1})\\
& \leq \min_{i\in \{1, \ldots, n \}} - \log \pi(i) \ud{}{1}(z_i, \ldots, z_n \mid z^{i-1}) \cdot \ud{}{0}(z^{i-1}) \\
& \leq  \min_i -\ \{  \log \ud{}{1}(z_i, \ldots, z_n \mid z^{i-1}) - \log  \ud{}{0}(z^{i-1}) \ \} \ + 2 \log n.
\end{align*}
In words, the cumulative log loss achieved by $\ud{\switch}{}$ is `almost' (within an order $\log n$ term) as small as that of
the strategy that first predicts by $\ud{}{0}$ and then {\em
  switches\/} from $\ud{}{0}$ to $\ud{}{1}$ at the switching point
$i$ that is optimal with hindsight. By clever choice of the prior
$\pi$, one can get the extra term down to order $\log \log n$. In
cases where the data are actually sampled from a distribution in
$\cM_1$ that is `close' (defined suitably) to $\cM_0$, the predictions
based on $\ud{\switch}{}$ will, with high probability, be substantially better
than those based on $\ud{}{1}$ --- a dramatic example (that makes very clear why this happens) is given in the first figure of \cite{erven2012catching}. If the data come from a distribution
that is `far' from $\cM_0$, they will tend to be worse than those based on
$\ud{}{1}$ by a negligible amount. Working out the math shows that
associating $\cM_1$ with $\ud{\switch}{}$ and $\cM_0$
with $\ud{}{0}$ indeed gives a strongly consistent model selection criterion that is almost (to within
an $O(\log \log n)$ factor) prediction optimal, thus almost solving
the AIC-BIC dilemma. \cite{erven2012catching} describes in great detail why the standard NML or Bayesian universal model $\ud{}{1}$ does not lead to the optimal cumulative log-loss if data come from a distribution close to, but not in, $\cM_0$. 

In case the number of models on the list is larger than two or even
infinite, one has to associate each model with a separate switch
distribution. The technique for doing so is described by \cite{erven2012catching} who also give an efficient implementation and prove consistency and
prediction optimality of the switch distribution in a weak, cumulative
sense for both finite and infinite numbers of models.  \cite{PasG17}
mathematically show the `almost' prediction optimality for a finite number of models.

\subsection{Hybrids Between NML, Bayes and Prequential Plug-In}

\paragraph{A Problem for NML: unknown horizon} Bayesian universal distributions with fixed priors have the property that the probability assigned to any initial sequence $z^{n'}$, where $n'<n$, is independent of the total length of the sequence. For other universal models, such as NML, this is not always the case. Take for example the Bernoulli model extended to sequences by independence: For sequence length $n=2$, the normalizing term in the NML equals $1+\left(1 /  2\right)^2+\left(1 / 2\right)^2+1 = 5/2$. For sequence length $n=3$, the normalizing term equals $2+6 \times \left(1 / 3\right)\left(2 /  3\right)^2 = 78/27$. For $n=2$, the NML probability of the sequence $00$ is $1/(5/2) = 0.4$. However, for sequence length is $n=3$, the probability of the initial sequence $00$ is obtained as the sum of the probabilities of the sequences $000$ and $001$, which becomes $1/(78/27) + (4/27)/(78/27) \approx 0.397 < 0.4$. As shown by  \cite{bartlett2013horizon}  (see also \cite{barron14}), there do exist cases in which NML is, like Bayes, horizon-independent, but these are very rare --- see Appendix~\ref{app:nml}. 

The above discrepancy between the initial sequence probabilities for different sequence lengths may be a problem in situations where we need to obtain predictions without necessarily knowing the total length of the sequence, or the \emph{horizon}. Another related issue is that even if the total sequence length was given, it can be computationally expensive to obtain marginal and conditional probabilities along the initial sequences.
One possible solution would be to restrict to Bayesian universal distributions. However, while these solve the horizon issue, they are (a) still often computationally inefficient, and (b) they lack NML-style worst-case regret interpretations. This has spurned research into universal codes that can be calculated  without knowing the horizon in advance and that behave better as regards to (a) or (b), which we now review.
\subsubsection{Prequential Plug-In and the \protect{$(k/2) \log n$}-formula} 
\label{sec:kotlowski}
The most straightforward way to deal with issue (a) is to use the prequential $\ud{\preq}{}$ which, by construction, is horizon-independent. However, 
for the prequential $\ud{\preq}{}$ (\ref{eq:prequential}) the BIC asymptotics (\ref{eq:kover2}) only hold in expectation if the data are sampled from one of the distributions in the model $\cM$. This makes the result much weaker than for the other five universal distributions considered, for which the asymptotics hold for every individual sequence in some large set, i.e. without making any stochastic assumptions at all.  One might thus wonder what happens for general data. Extending earlier work by \cite{TakeuchiB98b}, 
\cite{kotlowski2010following} show that, if data are sampled from a distribution $p$, and $p_{\tilde\theta}$ is the distribution in $\cM$ that is closest in KL divergence  to $p$, then (\ref{eq:kover2}) holds in expectation, with a correction term involving the variances of both distributions; for 1-dimensional models, we get
\begin{equation}\label{eq:corry}
\frac{\text{\sc var}_p(Z)}{\text{\sc var}_{p_{\tilde\theta}}(Z)} \cdot \frac{1}{2} \log n, 
\end{equation}
a formula that can be extended to multidimensional models and individual sequence settings. Solving Open Problem 2 from G07, \cite{GrunwaldK10} show that, essentially, there exists {\em no\/} `in-model' estimator that can achieve the standard asymptotics in general; a correction such as (\ref{eq:corry}) is always needed, whatever estimator one tries. Here an `in-model' estimator (or `proper' estimator, see end of Section~\ref{sec:codelength})  is an estimator that always outputs a distribution inside the model $\cM$; the ML and Bayes MAP estimators are in-model, but the Bayes predictive distribution is not in-model, since it is a mixture distribution over all distributions in $\cM$.

Solving Open Problem 3 from G07, \cite{kotlowski2010following}  also provide a new  universal distribution, in which for any given estimator $\breve\theta$, $\pud{\preq}{\breve\theta}{}(Z_{i+1} \mid z^{i}) = p_{\breve\theta(z^i)}(Z_{i+1})$ is turned into a slightly  `flattened' version $\pud{\preq *}{\breve\theta}{}(Z_{i+1} \mid z^{i})$, which is not in $\cM$ any more (it is not an in-model estimator), but it does achieve the standard $(k/2) \log n$ asymptotics without correction. For example, in case of the normal location family with fixed variance $\sigma^2$, it coincides with a Bayesian predictive distribution based on a standard conjugate prior, which in this case is a normal with mean $\breve\mu$ (the Bayes MAP estimate) but variance $\sigma^2 + O(1/n)$. More generally, $\pud{\preq *}{\breve\theta}{}(Z_{i+1} \mid z^{i})$ becomes a hybrid between the estimator $\breve\theta$ and a Bayes predictive distribution, but it has the advantage over the latter that it can be calculated without performing an integral over the parameter space. It thus provides an alternative to the NML distribution that is horizon-free and that is often faster to compute than $\ud{\bayes}{}$.

\cite{RoosRissanen2008} and \cite{RissanenRM10} developed other prequential, horizon-free universal codes that are non-Bayesian, yet remain more closely to NML in spirit than $\pud{\preq*}{\breve\theta}{}$. They work out the details for discrete models including Bernoulli as well as linear regression models. For Bernoulli models, the resulting universal code coincides with the so called one-step lookahead model proposed earlier by~\cite{takimoto00}. For linear regression models the asymptotic consistency of the resulting model selection criterion was studied by~\cite{maatta16a} and~\cite{maatta16b}.
Relatedly,  \cite{watanabe2015achievability} show that no horizon-independent strategy can be asymptotically minimax in the multinomial case and propose simple Bayesian universal models with a horizon-dependent Dirichlet prior that achieve asymptotic minimaxity and simplify earlier proposals. Among the proposed priors is $\mathrm{Dir}(\alpha,\ldots,\alpha)$ with $\alpha = {1/ 2} - {\ln 2 /  2\ln n}$ which converges to the Jeffreys' prior $\mathrm{Dir}({1 / 2},\ldots,{1/ 2})$ but has a mild dependency on the horizon $n$.

\subsection{Hypothesis Testing: Universal Distributions based on the Reverse Information Projection}
\label{sec:svalues}
Suppose we compare just two models, $\cM_0$ and $\cM_1$, as explanations for data $z^n$, a situation similar to classical {\em  null hypothesis testing}, the standard method for evaluating new treatments in the medical sciences and scientific hypotheses in most applied sciences such as psychology, biology and the like: we can think of $\cM_0$ and $\cM_1$ as two hypotheses, where, usually, $\cM_0$ represents the status quo (`treatment not effective', `coin unbiased'). In case\footnote{In this subsection we view, for notational convenience, the elements of $\cM_j$ as probability distributions $P_{\theta}$ with densities or mass functions $p_{\theta}$.} $\cM_0 = \{ P_0 \}$
represents a {\em simple\/} (singleton) hypothesis, there is a strong additional motivation for using MDL as a hypothesis testing method, and in particular, for quantifying the evidence against $\cM_0$ in terms of 
$$
D(z^n) = - \log \ud{}{1}(z^n)  - [ - \log p_0(z^n)  ],
$$
the codelength or cumulative-log-loss difference (see Section~\ref{sec:codelength}) between encoding (or sequentially predicting) the data with $p_0$ and with $\ud{}{1}$.  This additional motivation is given by the {\em no hyper-compression inequality\/} (G07), a mathematical result stating that, no matter how $\ud{}{1}$ is defined, as long as it is a probability distribution, we have for all $K > 0$, and $0 \leq \alpha \leq 1$
\begin{equation}\label{eq:nohyper}
P_0(D(Z^n) \leq - K) \leq 2^{-K}, \text{\ i.e.\ }
P_0 \left( \frac{p_0(Z^n)}{\ud{}{1}(Z^n)} \leq \alpha \right) \leq \alpha.
\end{equation}
This expresses, in terms of sequential log-loss prediction (compression), that, if $P_0$  is true, then the probability that one can predict data better, by $K$  or more loss units, by predictions based on  $\bar{p}_1$ rather than $p_0$, is exponentially small in $K$ --- and this holds independently of the sequence length $n$. In terms of more classical quantities, it states that, no matter how we chose $\ud{}{1}$, if $P_0$ holds true then the likelihood ratio is a $p$-value. In fact it is a conservative $p$-value, giving usually somewhat less evidence against $\cM_0$ than a standard $p$-value, for which the rightmost inequality in (\ref{eq:nohyper}) is an equality. The inequality (\ref{eq:nohyper}) goes in the right direction to retain the cornerstone of classical Neyman-Pearson testing: if one sets significance level $\alpha$ before seeing the data and one chooses $\cM_1$ whenever $D(Z^n) \leq - \log \alpha$, i.e. ${p_0(Z^n)}/{\ud{}{1}(Z^n)} \leq \alpha$, then the probability, under the null $P_0$, of making a false decision is bounded by $\alpha$. But the fact that the rightmost inequality in (\ref{eq:nohyper}) is usually strict has pleasant practical repercussions: as explored by \citep{GrunwaldHK19}, the Type-I error guarantees are retained under {\em optional continuation}. This is the (common) practice to decide, on the basis of an initial sample $z^n$, whether or not to gather new data and do a second test. One may for example decide to gather new data if the result based on $z^n$ was hopeful yet not conclusive. This is highly problematic for standard, strict $p$-value based hypothesis testing, but with MDL testing with a simple $\cM_0$, one can simply multiply the likelihood ratios of the two (or more) tests performed, or equivalently, add the code length differences for each test performed. The resulting code length difference/likelihood ratio will still lead to valid Type-I error bounds \citep{GrunwaldHK19}. 

But: all this holds only for simple  $\cM_0$. Yet the tests most used in practice, such as the $t$-test and contingency table tests, all involve {\em composite\/} $\cM_0 = \{ P_{\theta}: \theta \in \Theta_0 \}$. For composite $\cM_0$, the no-hypercompression inequality (\ref{eq:nohyper}) usually only holds for {\em some\/} $P_0 \in \cM_0$, but for Type I error guarantees and the like we would want to have it hold for {\em all\/} $P_{\theta}$ with $\theta \in \Theta_0$. That is, we would like to employ universal distributions $\ud{}{1}$ and $\ud{}{0}$ such that we have:
\begin{equation}\label{eq:nohyperb}
\text{For all $\theta \in \Theta_0$}: \ P_{\theta}\left(D(Z^n) \leq - K\right) \leq 2^{-K}, \text{\ i.e.\ }
P_{\theta}\left( \frac{\ud{}{0}(Z^n)}{\ud{}{1}(Z^n)} \leq \alpha\right) \leq \alpha.
\end{equation}
In general, this will not hold for standard choices (NML, Bayes, prequential plug-in...) of $\ud{}{1}$ and $\ud{}{0}$. However,  \cite{GrunwaldHK19} show that, for any given (arbitrary) $\ud{}{1}$, one can, under very mild conditions,  {\em construct\/} a $\ud{}{0}$ such that (\ref{eq:nohyperb}) holds, thereby solving Open Problems 9 and 19 of G07. This $\ud{}{0}$ is the {\em Reverse Information Projection\/} (RIPr) \citep{Li99,LiB00} of $\ud{}{1}$ onto ${\cal P}_{\bayes}(\cM_0)$, where ${\cal P}_{\bayes}$ is the set of densities $\ud{}{0}$ for $z^n$ that can be written as Bayes marginal distributions $\ud{\bayes}{0}(z^n) = \int p_{\theta} (z^n) w_0(\theta) d \theta$ for some prior $w_0$ on $\Theta_0$ --- for every prior $w_0$, ${\cal P}_{\bayes}(\cM_0)$ contains a separate distribution on $Z^n$. The RIPr is defined as the density achieving $\min_{\ud{}{0} \in  {\cal P}_{\bayes}(\cM_0) } 
D(\ud{}{1} \| \ud{}{0})$, where $D(\cdot \| \cdot)$ is the Kullback-Leibler divergence. Thus, one constructs a $\ud{}{0}$ with the desired no-hypercompression property, and at the same time, it will minimize KL divergence to $\ud{}{1}$, which implies that if data were sampled from $\ud{}{1}$, it would yield optimal log-loss predictions. This, in turn, implies that the $\ud{}{0}$ constructed this way will satisfy the standard asymptotics (\ref{eq:kover2}) as long as the $\ud{}{1}$ on which it is based does. Based on the likelihood ratio between $\ud{}{1}$ and its RIPr $\ud{}{0}$, one is also allowed to do optional continuation while retaining Type I Error guarantees. Thus, even if one is an adherent of classical, frequentist testing theory, there are  strong reasons for MDL-style testing based on the RIPr universal distribution. \cite{GrunwaldHK19} further extend the reasoning to give guidelines how $\ud{}{1}$ can be chosen to get further good frequentist properties.
\begin{example}{\bf\ [Right Haar Priors and the Bayesian $t$-test]}\label{ex:haar} In a series of papers (highlights include \citep{berger1998bayes,dass-2003-unified-condit}), Berger and collaborators established Bayes factor testing methods for composite $\cM_0 = \{ P_{\theta}: \theta \in \Theta_0\}$ where the only free parameters in $\Theta_0$ are `nuisance' parameters that are shared by $\Theta_1$ and are governed by a group structure. A prime example is the unknown variance in the $t$-test. Berger uses a special type of improper prior, the so-called {\em right-Haar\/} prior, which can be defined for every such type of nuisance parameter. While Bayes factors usually don't combine well with improper priors, the Bayes factors for group invariance parameters equipped with the right-Haar prior behave remarkably well. \cite{GrunwaldHK19} show that, even though the right Haar priors are usually improper, they can also be understood from a purely MDL perspective: if $\ud{\bayes}{1}$ and $\ud{\bayes}{0}$ are equipped with the right Haar prior on the nuisance parameters, and the prior on the additional parameters in $\ud{\bayes}{1}$ satisfies some additional requirements, then both $\ud{\bayes}{1}$ and $\ud{\bayes}{0}$ can be interpreted as sequential prediction strategies, and the log of the Bayes factor can be interpreted as the code length/cumulative log loss difference. Moreover, $\ud{\bayes}{0}$ is (essentially) the RIPr for $\ud{\bayes}{1}$ and the no-hypercompression inequality (\ref{eq:nohyperb}) that is so desirable from a frequentist perspective holds uniformly for all $\theta_0 \in \Theta_0$.

Let us consider the one-sample Bayesian $t$-test as an example. Here $\cM_0 = \{p_{0,\sigma} : \sigma > 0 \}$ is the set of all normal distributions with mean $0$; the variance $\sigma^2$ is a free parameter. $\cM_{1} = \{p_{\mu,\sigma} : \mu \in {\mathbb R}, \sigma > 0\}$ is the set of all normal distributions with as free parameters $\mu$ and $\sigma$. The question of interest is to establish whether $\mu = 0$ or not; $\sigma$ is an unknown `nuisance' parameter --- it determines the scale of the data but is not itself of intrinsic interest. In the {\em Bayesian $t$-test\/} one equips both $\cM_0$ and $\cM_1$ with the improper right-Haar prior, $w(\sigma) = 1/\sigma$. To complete the definition of $\ud{\bayes}{1}$, $\cM_1$ is equipped with a conditional prior density (given $\sigma$)  on the effect size $\delta := \mu/\sigma$. This second density has to be symmetric around $0$ and proper (this is what we called the `additional requirement on the prior on $\Theta_1$', instantiated to the case where the nuisance parameter is a variance). One now proceeds by testing using the Bayes factor $\ud{\bayes}{0}/\ud{\bayes}{1}$. In this special case, the procedure was already suggested by Jeffreys \citep{LiVW16}, and the right-Haar prior coincides with Jeffreys' prior on the variance. Berger et al. extend the method to general group invariant parameter vectors such as the joint mean and variance in the two-sample $t$-test, testing a Weibull against a log-normal and many other scenarios.
\end{example}
\section{Graphical Models}
\label{sec:graphical}
Graphical models are a framework for representing multivariate
probabilistic models in a way that encompasses a wide range of
well-known model families, such as Markov chains, Markov random
fields, and Bayesian networks; for a comprehensive overview,
see~\cite{Koller-book}. A key property of a graphical model is
parsimony, which can mean, for instance, a low-order Markov chain or
more generally a sparse dependency graph that encodes conditional
independence assumptions. Choosing the right level of parsimony in  graphical models is an ideal problem for MDL model selection.

In Bayesian network model selection the prevailing paradigm is,
unsurprisingly, the Bayesian one. Especially the work of~\cite{Geiger:1995:CDD:2074158.2074181,heckerman1995learning} has
been extremely influential. The main workhorse of this approach is the so called Bayesian Dirichlet (BD) family of scores which is applicable in the discrete case where the variables being modelled are categorical. Given a data sample, such scores assign a goodness value to each model structure. Exhaustive search for the highest scoring structure is possible when the problem instance (characterized by the number of random variables) is of limited size, but heuristic search techniques such as variants of local search or ``hill-climbing'' can be used for larger problem. 

Different forms of the BD score imply different Dirichlet priors (different hyper-parameters) for the local multinomial distributions that comprise the joint distribution. For
example, in the commonly used BDeu score, the priors are determined by
a single hyper-parameter, $\alpha$. For a variable $X_i$ with $r$
distinct values and parents $\mathrm{Pa}_i$ that
can take $q$ possible combinations of values (\emph{configurations}),
the BDeu prior is
$\mathrm{Dir}({\alpha /  rq},\ldots,{\alpha /  rq})$. One of the
main motivations for adopting this prior is that it leads
\emph{likelihood equivalence}, i.e., it assigns equal scores to all
network structures that encode the same conditional independence
assumptions. In light of the fact that Bayesian model selection embodies a particular form/variation of MDL, these method fit, at least to some extent, in the MDL framework as well. However, there also exist more `pure', non-Bayesian MDL methods for model selection in Bayesian networks; we mention \cite{LamBacchus1994} and \cite{Bouckaert2005} as early representative examples. These early methods are almost invariably based on the two-part coding framework. More recently, several studies have proposed new model selection criteria that exploit the NML distribution.
One approach is a continuous relaxation of NML-type complexities proposed by \cite{MiyaguchiMY17} in which the model selection problem takes on a tractable Lasso-type $L_1$-minimization form (see also Example~\ref{ex:varsel}). In other approaches, NML (or usually, approximations (but not relaxations) thereof) are used directly for encoding parts of the model; we now describe these latter approaches in a bit more detail. 

\subsection{Factorized NML and Variants}

\cite{ijar-paper} propose the \emph{factorized NML} (fNML) score for Bayesian network model selection which was designed to be \emph{decomposable} meaning that it can be expressed as a sum that includes a term for each variable in the network.  This property facilitates efficient search among the super-exponential number of possible model structures; see, e.g.,~\cite{heckerman1995learning}. The fNML score factors the joint likelihood not only in terms of the variables but also in terms of distinct configurations of the parent configurations. Each factor in the product is given by a multinomial NML probability, for which a linear-time algorithm by \cite{kontkanen2007linear} can be used.

A similar idea where a Bayesian network model selection criterion is constructed by piecing together multiple NML models under the multinomial model was proposed recently by \cite{Silander2018}. In the proposed \emph{quotient NML} (qNML) score, the local scores corresponding to each variable in the network are defined as log-quotients of the form
$$
\log  \frac{\mathrm{NML}_{\text{\sc full}}(X_i \cup \mathrm{Pa}_i) }{ \mathrm{NML}_{\text{\sc full}}(\mathrm{Pa}_i)},
$$ 
where $\mathrm{NML}_{\text{\sc full}}$ refers to an NML distribution defined by using a fully connected network to model the variable $X_i$ and its parents $\mathrm{Pa}_i$ in the numerator and the same thing for the parent set $\mathrm{Pa}_i$ in the denominator. Technically, this amounts to collapsing the configurations of the variables into distinct values of a single categorical variable. Even though the resulting categorical variable may have a huge number of possible values, the linear time algorithm~\citep{kontkanen2007linear} or efficient approximations (see the next subsection) can be used to implement the computations. A notable property of the qNML score is that, unlike the fNML score, it is likelihood equivalent (see above).

\cite{Eggeling14}
apply similar ideas to a different model class, namely parsimonious Markov chains. There too, the likelihood is decomposed into factors depending on the configurations of other variables, and each part in the partitioning is modelled independently using the multinomial NML formula. The authors demonstrate that the fNML-style criterion they propose leads to parsimonious models with good predictive accuracy for a wide range of different scenarios, whereas the corresponding Bayesian scores are sensitive to the choice of the prior hyperparameters, which is important in the application where parsimonious Markov chains are used to model DNA binding sites~\citep{Eggeling15}.

In all these papers, both simulated and real-world data experiments suggest that the MDL-based criteria are quite robust with respect to the parameters in the underlying data source. In particular, the commonly used Bayesian methods (such as the BDeu criterion) that are being used as benchmarks are much more sensitive and fail when the assumed prior is a poor match to the data-generating model, whereas the MDL methods are invariably very close to the Bayesian methods with the prior adapted to fit the data. This poses interesting questions concerning the proper choice of priors in the Bayesian paradigm.

In fact, the prevalence of the Bayesian paradigm and the commonly used BD scores is challenged by two recent observations: First, \cite{ijar-paper} show that the Dirichlet prior with hyperparameters $(1/2,\ldots,1/2)$, which is the invariant Jeffreys prior for the multinomial model, but not likelihood equivalent when used in the BD score, is very close to the fNML model and consequently, enjoys better robustness properties than the BDeu score which is the likelihood equivalent BD score variant. Second, \cite{profJoe-witmse2016} shows that the BDeu criterion is \emph{irregular}, i.e., prone to extreme overfitting behavior in situations where a deterministic relationship between one variable and a set of other variables holds in the data sample. The MDL scores discussed above are regular in this respect and their robustness properties seem to be better than those of the BD scores, see~\citep{Silander2018}.

\subsection{Asymptotic Expansions for Graphical Models}\label{sec:graphicalasymptotics}

Asymptotic results concerning MDL-based criteria in graphical models are interesting for several reasons. For one, they lead to efficient scores that can be evaluated for thousands of different model structures. Secondly, asymptotic expansions can lead to insights about the relative complexity of different model structures.

Various asymptotic forms exist for the point-wise and the expected regret depending on the model class in question. For convenience we repeat the classical expansion of the NML (as well as the Bayesian marginal likelihood with Jeffreys' prior) regret/model complexity that applies for regular model classes $\cM = \{ p_{\theta}: \theta \in \Theta \}$ for which $\comp(\cM,v)$ is finite with uniform $v$ (see Section~\ref{sec:asymptotic} above):
%, which agrees with Eq.~\ref{eq:nmlasymptotics} when the luckiness function is a constant: 
\begin{equation}\label{eq:classical}
\comp(\cM,v) =  {k\over 2}\log {n\over 2\pi} + \int_\Theta \sqrt{|I(\theta)|} d\theta + o(1),
\end{equation}
where $k$ is the dimension of the model, $|I(\theta)|$ is the determinant of the Fisher information matrix at parameter $\theta$, the integral is over the parameter space $\Theta$, and the remainder term $o(1)$ vanishes as the sample size tends to infinity.

For discrete data scenarios, by far the most interesting case is the multinomial model (extension of the Bernoulli distribution to an i.i.d.\ sequence of $r$-valued categorical random variables) since it is a building block of a number of MDL-criteria such as fNML and qNML (see above). There are many asymptotic expansions for the NML regret under the multinomial model. Probably the most useful is the one proposed by~\cite{Szpankowski2012}:
\begin{equation}\label{eq:szpwein}
n\left(\log \alpha + (\alpha+2)\log C_\alpha - {1\over C_\alpha}\right)
-{1\over 2} \log\left(C_\alpha + {2\over \alpha}\right),
\end{equation}
where $n$ is the sample size, $\alpha={r\over n}$, and $C_\alpha={1\over 2} + {1\over 2}\sqrt{1+{4\over \alpha}}$. This simple formula is remarkably accurate over a wide range of finite values of $n$ and $r$ (see~\cite{Silander2018}). Note that the leading term is proportional to $n$ (rather than $\log n$ as usual) because the formula is derived for the regime $r = \Theta(n)$ where the alphabet size grows proportionally to the sample size. If $r$ grows slower than $n$ or not at all, the leading term tends to the classical form~(\ref{eq:classical}), where the leading term is ${k\over 2}\log n$. In practice, the approximation~(\ref{eq:szpwein}) is applicable for a wide range of $r/n$ ratios.

\cite{itw2008} and~\cite{ngc2017} studied the second term in the expansion  (\ref{eq:classical}), namely the Fisher information integral, under Markov chains and Bayesian networks using Monte Carlo sampling techniques. This approach reveals systematic differences between the complexities of models even if they have the same number of parameters.

\section{Latent Variable and Irregular Models}
\label{sec:latent}
Although thus far we have highlighted exponential family and regression applications, NML and other universal distributions can of course be used for model selection and estimation in complete generality --- and many practical applications are in fact based on highly irregular models. Often, `classical' 2-part distributions (based on discretized models) 
are used, since NML distributions often pose computational difficulties. However, Yamanishi and collaborators have managed to come up with tractacble approximations of NML-type distributions for some of the most important irregular (i.e. non-exponential family) models such as  hierarchical latent variable models \citep{WuSY17}, and the related  Gaussian mixture models \cite{HiraiY17,hirai2013efficient}. 
\cite{suzuki2016structure} provides an NML approach to nonnegative matrix factorization. Two-part codes (and corresponding MDL estimators) for mixture families that come close to achieving the minimax regret were considered very recently by \cite{MyamotoBT19}.

When it comes to asymptotic approximations for codelengths/log-likelihoods based on NML and other universal distributions --- all approximations so far (in Section~\ref{sec:asymptotic} were  derived essentially assuming that the model under consideration is an exponential family. Extensions to curved exponential families and generalized linear models are relatively straightforward (see G07 for details).  
For more irregular models, Sumio Watanabe has proposed the widely applicable information criterion (WAIC) and the widely applicable Bayesian information criterion (WBIC), see~\citep{watanabe2010,watanabe2013}, where the latter can be viewed as an asymptotic expansion of the log likelihood based on a Bayesian universal distribution. It coincides with BIC when applied to regular models but is applicable even for singular (irregular) models. The asymptotic form of WBIC is
\begin{equation}
\mathrm{WBIC}(\cM) = -\log p_{\theta_0}(z^n) + \lambda \log n + O_p(\sqrt{\log n}),
\end{equation}
where $\theta_0$ is the parameter value minimizing the Kullback-Leibler divergence from the model to the true underlying distribution,  and $\lambda>0$ is a rational number called the real log canonical threshold (see~\citep{watanabe2013}), which can be interpreted as the effective number of parameters (times two).
\section{Frequentist Convergence of MDL and Its Implications}\label{sec:freq}
Rissanen first formulated the MDL Principle as --- indeed --- a {\em Principle\/}: one can simply start by {\em assuming}, as an axiom, that model  ing by data compression (or, equivalently, sequential predictive log loss minimization) is the right thing to do. One can also take a more conventional, frequentist approach, and check whether MDL procedures behave desirably under standard frequentist assumptions. We now review results that show that, in general, they do --- thus providing a frequentist justification of MDL ideas: with some interesting caveats, MDL model selection is typically {\em consistent\/} (the smallest model containing the true distribution is eventually chosen, with probability one) and MDL prediction and estimation achieves good {\em rates of convergence} (the Hellinger distance between the estimated and the true density goes to 0, with high probability, quite {\em fast\/}). In this section we review the most important convergence results. In particular, Section~\ref{sec:mehta} shows that the link between data compression and consistent estimation is in fact very strong;  and Section~\ref{sec:deep} shows that, by taking MDL as a {\em principle}, one can get useful intuitions about deep questions concerning deep learning; and the intuitions can then, as a second step, be once again validated by frequentist results.  

Thus, let us assume, as is standard in frequentist statistics, that data are drawn from a distribution in one of the models under $\cM_\gamma$ under consideration. We consider consistency and convergence properties of the main MDL procedures in their main applications: model selection, prediction and estimation.
\paragraph{Model Selection}
For model selection between a finite number of models, all universal codes mentioned here are consistent in wide generality; for example, this has been explicitly proven if the data are i.i.d. and all models on the list are exponential families, but results for more complex models with dependent data have also been known for a long time; see G07 for an overview of results. If the collection of models is countably infinite, then results based on associating each $\cM_{\gamma}$ with $\ud{\bayes}{ \gamma}$ have also been known for a long time; such results typically hold for `almost all' (suitable defined) distributions in all $\cM_{\gamma}$; again, see G07 for a discussion of the (nontrivial) `almost all' requirement. These countable-$\Gamma$ consistency results were extended to the switch distribution by \cite{erven2012catching}. 

\paragraph{Prediction and `Improper' Estimation}
As to sequential prediction (Section~\ref{sec:codelength}): rate of convergence results are very easy to show (see Chapter 15 of G07), but these typically only demonstrate that the {\em cumulative\/} log-loss prediction error of sequentially predicting with a universal distribution $\ud{}{}$ behaves well as $n$ increases. Thus, since the sum of prediction errors is small, say (for parametric models) of order $\log n$, for  {\em most\/} $t$ the  individual prediction error at the $t$-th sample point must be of order $1/t$, since $\sum_{t=1}^n 1/t - \log t = O(1)$. Still,  it remains an open question how to prove for individual $t$ what exactly the expected prediction error is at that specific $n$. Since one can view each prediction as an `improper' estimate (end of Section~\ref{sec:codelength}), the convergence rates of the resulting estimators, which estimate the underlying distribution based on a sample of size $t$ as $\ud{}{}(Z_{t+1} \mid z^t)$, usually also behave well in a cumulative sense, but again it is very hard to say anything about individual $t$. The asymptotic expansions (\ref{eq:nmlasymptotics}) and (\ref{eq:bayesasymptotics}) imply that, for fixed parametric models $\cM_{\gamma}$, $\ud{\bayes}{\gamma}$ and $\ud{\nml}{\gamma}$ achieve optimal cumulative prediction and estimation errors. If, however, they are defined relative to a full model class  $\cM = \bigcup_{\gamma \in \Gamma} \cM_{\gamma}$ consisting of at least two nested models, then they may fail to achieve optimal rates by a $\log n$ factor. \cite{erven2012catching} show that sequential prediction/estimation based on the switch distribution achieves the minimax optimal cumulative prediction/estimation error rates even in such cases. \cite{PasG17} show that, if only two models are compared, then the optimal obtainable rate for individual $n$ for any consistent procedure is achieved as well.  

\subsection{Frequentist Convergence of MDL Estimation}
\label{sec:mehta}
Very strong results exist concerning the convergence of MDL estimation based on an MDL estimator $\mdlest{\theta}{v}$ as given by (\ref{eq:brevepre}). A first, classical result was already stated by the ground-breaking \citep{BarronC91}, establishing that consistency and good convergence rates can be obtained for the special case of a two-part-code estimator $\mdlest{\theta}{w}$ based on a probability mass function $w$, as long as $w$ satisfies $\sum_{\theta \in \ddot{\Theta}} w(\theta)^{\eta} < \infty$ for some  $\eta < 1$ and $w$ puts sufficient prior mass in a KL neighborhood of the true $\theta$. These results were greatly extended by \cite{Zhang06a,Zhang06b} and further, very recently, by \cite{GrunwaldM19}. The latter consider $\mdlest{\theta}{v}$ for general $v$. Let $\cM = \{P_{\theta} : \theta \in \Theta \}$ be a statistical model and suppose data are i.i.d. $\sim p$ with $p = p_{\theta^*} \in \cM$. They find that a sufficient condition for consistency is that $v(\theta^*) < \infty$ and that for some $\eta < 1$, the following {\em generalized model complexity\/}
\begin{equation}\label{eq:complexityb}
\comp_{\eta}(\cM,v) := \comp(\cM_{\eta},v) =  \log \int p(z^n)^{1-\eta} \cdot 
\frac{\left(  p_{\mdlest{\theta}{v}}(z^n)\right)^{\eta} v(\mdlest{\theta}{v}(z^n))} {\left(\int p(\underline{z})^{1-\eta} \left(p_{\mdlest{\theta}{v}(z^n)}(\underline{z})\right)^{\eta}  
d \underline{z} \right)^n}
d z^n
\end{equation}
is bounded and $o(n)$, i.e. it grows slower than linear (the slower it grows, the faster the MDL estimator converges to the true distribution in Hellinger distance). This condition strictly and significantly weakens the Barron-Cover requirement.  The result holds without any further conditions; for example, $\cM$ may be a countable union of parametric models or even a huge nonparametric model. Note that $\comp_1(\cM,v)$ is the model complexity that we have encountered before in (\ref{eq:complexity}). Ironically, for any $\eta < 1$, slow-growth ($o(n)$) of $\comp_{\eta}(\cM)$ is sufficient for consistency of $\mdlest{\theta}{v}$, but for $\eta=1$, which would be more fully in line with the MDL ideas, it is not. 
\subsection{From MDL to Lasso}\label{sec:lasso}
As illustrated in Example~\ref{ex:varsel}, when used for high-dimensional variable selection, the original MDL approach would be to use a mixed two-part/one-part code as in (\ref{eq:selectingk}) with a $- \log \pi(\gamma)$ term to account for the model index $\gamma \in \Gamma$. In such settings, there may well be $p >  n$ variables of interest, each of which may or may not be included in the model, so that the minimization over $\Gamma$ requires trying out $2^p \gg 2^n$ choices --- which is practically infeasible. For this reason, in practice people have strongly preferred the {\em Lasso\/} and related methods based on $L_1$-penalties, which take linear rather than exponential time in $p$ (note that the classic MDL essentially penalizes by an $L_0$-penalty). However, \cite{BarronL08,BarronHLL08} showed that, under some conditions on the true distribution (such as Gaussian noise), the Lasso method can be re-interpreted in terms of code-length minimization after all; see also \citep{ChatterjeeB14}, and, for further extensions,  \cite{KawakitaT16,BrindaK18}. For a different approach to unify model selection with very high-dimensional models with the luckiness NML, see \cite{MiyaguchiY18}. 

Although some of the details may differ, it seems that most of these works are subsumed by the aforementioned result of \cite{GrunwaldM19}
who show that general penalized estimators can be re-interpreted as minimizing a 1-part codelength as long as $\comp_1(\cM,v)$ is bounded, and can be proven consistent under the (still quite weak) condition that $\comp_{\eta}$ as in (\ref{eq:complexityb}) is bounded for some $\eta < 1$. Thus, the connection between MDL and general (including Lasso and other $L_1$, but also with entirely different penalties) penalization methods is substantially stronger than it seemed before these developments took place. 

\paragraph{Supervised Machine Learning} Importantly, all the works mentioned here except \cite{GrunwaldM19} cannot show convergence under misspecification --- for example, when applied to the Lasso, they would require an assumption of normal noise (corresponding to the squared error used in the Lasso fit, which is equivalent to the log loss under a normal distribution for the noise). In practice though, the Lasso (with the squared error) is often used in cases in which one cannot assume normally distributed errors. \cite{GrunwaldM19} contains results that can still be used in such cases (although the formula for $\comp_{\eta}(\cM,v)$ changes), based on ideas which we sketch in the following section. 

More generally, one of the major areas within machine learning is {\em supervised learning\/} in which one assumes that data $(X_1,Y_1), (X_2,Y_2), \ldots$ are i.i.d. $\sim P_0$, with $X_i \in {\cal X}$ and $Y_i \in {\cal Y}$, and one aims to use the data to learn a predictor function $f: {\cal X} \rightarrow {\cal Y}'$ that has small {\em expected loss\/} or {\em risk}, defined as ${\bf E}_{(X,Y) \sim P} \left[{\ell}(f(X),Y)\right]$, where $\ell: {\cal Y}\times {\cal Y}' \rightarrow {\mathbb R}$ is some loss function of interest and $f$ is a member of some `predictor model' ${\cal F}$. For example, the statistical notion of `regression with random design' corresponds, in machine learning, to a supervised learning problem with ${\cal Y } = {\cal Y}' = {\mathbb R}$ and $\ell(y',y) = (y'-y)^2$. Early MDL convergence results do not cover this `supervised' situation: they are not equipped to handle either random design or loss functions beyond the log-loss. Some of the more recent works mentioned above are  able to handle random design but not general loss functions (for example, for Lasso-type applications they require the noise to be normally distributed). \cite{GrunwaldM19} seem to be the first that can fully handle supervised learning scenarios: the convergence results can be used with random design, and they can also be used with large classes of loss functions including squared error (without normality assumption) and $0/1$-loss. This is achieved by associating predictors $f$ with densities $p_f(x,y) \propto \exp(- \ell(f(x),y))$, so that the log-loss relative to density $p_f$ on data $(x,y)$ becomes linearly related to the loss of $f$ on $(x,y)$; the analysis then proceeds via analyzing convergence of MDL for the densities $\{ p_f: f \in {\cal F} \}$ as a misspecified probability model. 
\subsection{Misspecification}
\label{sec:misspecification}
As beautifully explained by \cite{Rissanen91}, one of the main original motivations for MDL-type methods is that they have a clear interpretation independent of whether any of the models under consideration is `true' in the sense that it generates the data: one chooses a model minimizing a code length, i.e. a prediction error on unseen data, which is meaningful and presumably might give something useful irrespective of whether the model is true (Rissanen even argues that the whole notion of a `true model' is misguided). This model-free paradigm also leads one to define the NML distribution as minimizing prediction error in a stringent worst-case-over-all data sense (Eq.~(\ref{eq:maximin}))  rather than a stochastic sense. Nevertheless, it is of interest to see what happens if one samples data from a distribution for which all models under consideration are wrong, but some are quite useful in the sense that they lead to pretty good predictions. Doing this leads to rather unpleasant surprises: as first noted by \cite{GrunwaldL07}, MDL (and Bayesian inference) can become inconsistent: one can give examples of $\{ \cM_{\gamma} : \gamma \in \Gamma \}$ with countably infinite $\Gamma$ and a 'true' data generating distribution $P_0$ such that, when data are sampled i.i.d. from $P_0$, MDL will tend to select a suboptimal model  for all large $n$ ---  while all sub-models $\cM_{\gamma}$ are wrong, one of them, $\cM_{\tilde\gamma}$ is optimal in several intuitive respects (closest in KL divergence to $P_0$, leading to best predictions under a number of loss functions), yet it will not be selected for large $n$. While the models considered by \cite{GrunwaldL07} were quite artificial, \cite{GrunwaldO17} showed that the same can happen in a more natural linear regression setting; moreover, they also showed that even if $\Gamma$ is finite, although then  {\em eventually\/} MDL will select the best sub-model, for even relatively large $n$ it may select arbitrarily bad sub-models.      \cite{de2016safe} shows that the problem also occurs with MDL and Bayesian regression with some real-world data sets. 

It turns out that the root of the problem is related to the no-hypercompression property (\ref{eq:nohyper}). If the collection of models $\cM = \bigcup_{\gamma \in \Gamma} \cM_{\gamma}$ contain the density $p_0$ of the `true' distribution $P_0$, then any distribution $p \in \bigcup_{\gamma \in \Gamma} \cM_{\gamma}$  will satisfy no-hypercompression relative to the true $p_0$: 
\begin{equation}\label{eq:nohyperc}
P_0 \left( \frac{p_0(Z^n)}{{p}(Z^n)} \leq \alpha \right) \leq \alpha.
\end{equation}
This property underlies the proof of all MDL consistency and rate-of-convergence results, such as those by \cite{BarronC91,Zhang06a,GrunwaldM19}.
However, if the model class $\cM$ does not contain the true $p_0$, then, in order to prove consistency, one needs (\ref{eq:nohyperc}) to hold with the $P_0$ outside the brackets unchanged, but the $p_0$ inside the brackets replaced by $\tilde{p}$, the distribution/density in $\cM$ that is closest to $P_0$ in KL (Kullback-Leibler) divergence (why it should be KL is explained at length by \cite{GrunwaldO17}). Unfortunately though, (\ref{eq:nohyperc}) does not necessarily hold with $p_0$ replaced by $\tilde{p}$. If it does not, MDL (and Bayesian methods, whose consistency relies on similar properties) may become inconsistent. \cite{GrunwaldO17}, based on earlier ideas in  \citep{Grunwald99a,Grunwald12}, propose a solution that works for Bayesian universal distributions: it replaces the likelihoods $p_{\theta}(z^n)$ for every $p = p_{\theta}$ with $p \in \cM$ by the {\em generalized likelihood\/} $p_{\theta}^{\eta}(z^n)$ for some $\eta > 0$; usually $\eta < 1$ --- this $\eta$ has the same mathematical function as the $\eta$ appearing in (\ref{eq:complexityb}). It turns out that with such a modification, if $\eta$ is chosen small enough, a version of the no-hypercompression inequality (\ref{eq:nohyperc}) holds after all. \cite{GrunwaldO17,Grunwald12} also provide a method for learning $\eta$ from the data, the `Safe Bayesian' algorithm (note that $\eta$ cannot be learned from the data  by standard MDL or Bayesian methods). The recent work of \cite{GrunwaldM19} suggests that the modification of likelihoods by exponentiating with $\eta$ should work for general MDL methods as well.    
\subsection{PAC-MDL Bounds and Deep Learning}
\label{sec:deep}
One of the great mysteries of modern {\em deep learning\/} methods in machine learning is the following \citep{zhou2018compressibility}: deep learning is based on {\em neural network\/} models which can have many millions of parameters. Although typically run on very large training samples $z^n$, $n$ is usally still so small that the data can be fit perfectly, with 0 error on the training set. Still, the trained models often perform very well on future test sets of data. How is this possible? At first sight this contradicts the tenet, shared by MDL and just about any other method of statistics, that good generalization requires the models to be `small' or `simple' (small $\comp(\cM)$ in MDL analyses, small VC dimension or small entropy numbers in statistical learning analyses) relative to the sample size. One of several explanations (which presumably all form a piece of the puzzle) is that the local minimum of the error function  found by the training method is often very {\em broad} --- if one moves around in parameter space near the minimum, the fit hardly changes. 
Hochreiter and Schmidhuber (\citeyear{HochreiterS97}) already observed that describing weights in sharp minima requires high precision in order to not incur nontrivial excess error on the data, whereas flat minima can be described with substantially lower precision, thus forging a
connection to the MDL idea; in fact related ideas already appear in
\citep{hinton1993keeping}. In these papers, the MDL Principle is used in a manner that is less direct than what was done thus far in this paper: we (and, usually, Barron, and Rissanen) {\em directly\/} hunt for the shortest description of the data. In contrast, the aforementioned authors simply note that, {\em no matter how} a vector of parameters for a model was obtained, if, with the obtained vector of parameters, the data can be compressed substantially, for example by coding first the parameters and then the data with the help of the parameters, then, if we believe the MDL principle, with these parameters the model (network) should generalize well to future data. In modern practice, neural networks are often trained with SGD (stochastic gradient descent), and it has been empirically found that networks that generalize well do tend to have parameters lying in very flat minima. 

While this use of the MDL Principle seems less precise than what we reviewed earlier in this paper, it can once again be given a frequentist justification, and this justification is mathematically precise after all: the so-called {\em PAC-Bayesian generalization bounds\/} \citep{McAllester02} show that the generalization performance of any classifier can be directly linked to a quantity that gets smaller as soon as one needs (a) less bits to describe the parameter and as soon as one needs (b) less bits to describe the data given the parameters; both the results and their proofs are very similar to the MDL convergence results by \cite{BarronC91,Zhang06a,Zhang06b,GrunwaldM19}. Although in general the formulation is not as straightforward as a simple sum of the two description lengths (a) and (b), the connections between both the two-part codelength and the Bayesian codelength are quite strong, as was already noticed by 
\cite{BlumL03}. In particular, for discrete $\Theta$, such PAC-Bayes bounds contain a term $- \log \pi(\theta)$ which can be interpreted as the number of bits needed to encode $\theta$ using the codes based on some distribution $\pi$; for general, uncountable $\Theta$, this term gets replaced by a KL divergence term that canstill be related to a codelength via a so-called `bits back argument' pioneered by \cite{hinton1993keeping}. \cite{dziugaite2017computing,zhou2018compressibility}, inspired
by earlier work by \cite{LangfordC02}, indeed show that, for some real-world data sets, one can predict nontrivial generalization using deep neural nets by looking at the number of bits needed to describe the parameters and applying PAC-Bayesian bounds.

\section{Concluding Remarks}\label{sec:concluding}
We have given a self-contained introduction to MDL, incorporating and highlighting recent developments. Of necessity, we had to make a choice as to what  to cover in detail, and there are many things we omitted. We would like to end with briefly mentioning three additional developments. First, there has always been the question about how MDL relates to other complexity notions such as those considered in the {\em statistical learning theory\/} literature \citep{shalev2014understanding}: Vapnik-Chervonkis dimension, entropy numbers, Rademacher complexity and so on. A major step towards understanding the relation was made by \cite{GrunwaldM19} who show that for probability models with members of  the form $p_{\theta}(z) \propto \exp(- \eta \text{\sc loss}_{\theta}(z))$, where $\text{\sc loss}$ is an arbitrary bounded loss function, the NML complexity can be precisely bounded in terms of the Rademacher complexity defined relative to $\text{\sc loss}$. Second, we should note that Rissanen's own views and research agenda have steered in a direction somewhat different from the developments we describe: 
\cite{Rissanen07} published {\em Information and Complexity in Statistical Modeling}, which proposes foundations of statistics in which no underlying `true model' is ever assumed to exist. As Rissanen writes, {\em ``even such a well-meaning statement as `all models are wrong, but some are useful', is meaningless unless some model is `true'. ''}  Rissanen expands MDL and NML ideas in the direction of the {\em Kolmogorov structure function}, taking the idea of distinguishable distributions underlying \cite{MyungBP00} as fundamental; while presumably compatible with the developments we describe here, the emphasis of this work is quite different.  

We end with a word about applications: since 2007,  numerous applications of MDL and MDL-like techniques have been described in the literature; as discussed in Section~\ref{sec:lasso}, highly popular methods such as Lasso and Bayes factor methods can often be seen as `MDL-like'. Even as to specific `pure' MDL applications (such as based on NML and two-part codes), the number and scope of applications is simply too large to give a succinct representative overview. However, there is one particular area which we would like to mention specifically, since that area had hardly seen any MDL applications before 2007 whereas nowadays such applications are flourishing: this is the field of {\em data mining}. Some representative publications are  
\cite{VreekenLS11,koutra2015summarizing,BudhathokiVO18}. Most of this work centers on the use of two-part codes, 
but sometimes NML and other sophisticated universal distributions/codes are used as well \citep{tatti2008finding}.
\section*{Acknowledgements}
We would like to thank Kenji Yamanishi, Matthijs van Leeuwen, and Bin Yu for providing references to new MDL work, Jun-Ichi Takeuchi for prompting us to write this paper, and Jorma Rissanen and Andrew Barron for many fruitful conversations over the last 20 years. This work is part of the research programme {\em Safe Bayesian Inference\/} with project number  617.001.651, which is financed by the Dutch Research Council (NWO). This work has also
been supported in part by the Academy of Finland (grant numbers 311277 and 311808).
\DeclareRobustCommand{\VANDER}[3]{#3}
\bibliographystyle{plainnat}
\bibliography{MDL}

\begin{thebibliography}{96}
\providecommand{\natexlab}[1]{#1}
\providecommand{\url}[1]{\texttt{#1}}
\expandafter\ifx\csname urlstyle\endcsname\relax
  \providecommand{\doi}[1]{doi: #1}\else
  \providecommand{\doi}{doi: \begingroup \urlstyle{rm}\Url}\fi

\bibitem[Bar-Lev et~al.(2010)Bar-Lev, Bshouty, Gr{\"u}nwald, and
  Harremo{\"e}s]{bar2010jeffreys}
Shaul~K Bar-Lev, Daoud Bshouty, Peter Gr{\"u}nwald, and Peter Harremo{\"e}s.
\newblock Jeffreys versus {S}htarkov distributions associated with some natural
  exponential families.
\newblock \emph{Statistical Methodology}, 7\penalty0 (6):\penalty0 638--643,
  2010.

\bibitem[Barron et~al.(1998)Barron, Rissanen, and Yu]{BarronRY98}
A.~Barron, J.~Rissanen, and B.~Yu.
\newblock The minimum description length principle in coding and modeling.
\newblock \emph{IEEE Transactions on Information Theory}, 44\penalty0
  (6):\penalty0 2743--2760, 1998.
\newblock Special Commemorative Issue: Information Theory: 1948-1998.

\bibitem[Barron et~al.(2008)Barron, Huang, Li, and Luo]{BarronHLL08}
A~Barron, Cong Huang, J~Li, and Xi~Luo.
\newblock The {MDL} principle, penalized likelihoods, and statistical risk.
\newblock In \emph{Festschrift in Honor of Jorma Rissanen on the Occasion of
  his 75th Birthday}, 2008.

\bibitem[Barron and Luo(2008)]{BarronL08}
Andrew Barron and Xi~Luo.
\newblock {MDL} procedures with {L}1 penalty and their statistical risk.
\newblock In \emph{Proceedings of the 2008 workshop on information theoretic
  methods in science and engineering}, 2008.

\bibitem[Barron et~al.(2014)Barron, Roos, and Watanabe]{barron14}
Andrew Barron, Teemu Roos, and Kazuho Watanabe.
\newblock Bayesian properties of normalized maximum likelihood and its fast
  computation.
\newblock In \emph{Proceedings of the IEEE International Symposium on
  Information Theory (ISIT-2014)}, pages 1667--1671. IEEE Press, 2014.

\bibitem[Barron and Cover(1991)]{BarronC91}
A.R. Barron and T.M. Cover.
\newblock Minimum complexity density estimation.
\newblock \emph{IEEE Transactions on Information Theory}, 37\penalty0
  (4):\penalty0 1034--1054, 1991.

\bibitem[Bartlett et~al.(2013)Bartlett, Gr{\"u}nwald, Harremo{\"e}s, Hedayati,
  and Kotlowski]{bartlett2013horizon}
Peter Bartlett, Peter Gr{\"u}nwald, Peter Harremo{\"e}s, Fares Hedayati, and
  Wojciech Kotlowski.
\newblock Horizon-independent optimal prediction with log-loss in exponential
  families.
\newblock In \emph{Proceedings 26th Conference on Learning Theory (COLT 2013)},
  2013.

\bibitem[Berger et~al.(1998)Berger, Pericchi, and Varshavsky]{berger1998bayes}
James~O Berger, Luis~R Pericchi, and Julia~A Varshavsky.
\newblock Bayes factors and marginal distributions in invariant situations.
\newblock \emph{Sankhy{\=a}: The Indian Journal of Statistics, Series A}, pages
  307--321, 1998.

\bibitem[Blum and Langford(2003)]{BlumL03}
A.~Blum and J.~Langford.
\newblock {PAC}-{MDL} bounds.
\newblock In \emph{Proceedings of the Sixteenth Conference on Learning Theory
  (COLT' 03)}, pages 344--357, 2003.

\bibitem[Bouckaert(2005)]{Bouckaert2005}
Remco~R. Bouckaert.
\newblock Probabilistic network construction using the minimum description
  length principle.
\newblock In M.~Clarke, R.~Kruse, and S.~Moral, editors, \emph{Symbolic and
  Quantitative Approaches to Reasoning and Uncertainty}, volume 747 of
  \emph{Lecture Notes in Computer Science}, pages 41--48. Springer, 2005.

\bibitem[Brinda and Klusowski(2018)]{BrindaK18}
W.~Brinda and J.~Klusowski.
\newblock Finite-sample risk bounds for maximum likelihood estimation with
  arbitrary penalties.
\newblock \emph{IEEE Transactions on Information Theory}, 2018.

\bibitem[Budhathoki et~al.(2018)Budhathoki, Vreeken, and Origo]{BudhathokiVO18}
K.~Budhathoki, J.~Vreeken, and J.~Origo.
\newblock Causal inference by compression.
\newblock \emph{Knowledge and Information Systems}, 56\penalty0 (2):\penalty0
  285--307, 2018.

\bibitem[Cesa-Bianchi and Lugosi(2006)]{CesaBianchiL06}
N.~Cesa-Bianchi and G.~Lugosi.
\newblock \emph{Prediction, Learning and Games}.
\newblock Cambridge University Press, Cambridge, UK, 2006.

\bibitem[Chatterjee and Barron(2014)]{ChatterjeeB14}
Sabyasachi Chatterjee and Andrew Barron.
\newblock Information theoretic validity of penalized likelihood.
\newblock In \emph{Information Theory (ISIT), 2014 IEEE International Symposium
  on}, pages 3027--3031. IEEE, 2014.

\bibitem[Cover and Thomas(1991)]{CoverT91}
T.M. Cover and J.A. Thomas.
\newblock \emph{Elements of Information Theory}.
\newblock Wiley-Interscience, New York, 1991.

\bibitem[Dass and Berger(2003)]{dass-2003-unified-condit}
Sarat~C. Dass and James~O. Berger.
\newblock Unified conditional frequentist and {B}ayesian testing of composite
  hypotheses.
\newblock \emph{Scandinavian Journal of Statistics}, 30\penalty0 (1):\penalty0
  193–210, Mar 2003.

\bibitem[Dawid(2007)]{dawid2007geometry}
A~Philip Dawid.
\newblock The geometry of proper scoring rules.
\newblock \emph{Annals of the Institute of Statistical Mathematics},
  59\penalty0 (1):\penalty0 77--93, 2007.

\bibitem[Dawid(1984)]{Dawid84}
A.P. Dawid.
\newblock Present position and potential developments: Some personal views,
  statistical theory, the prequential approach.
\newblock \emph{Journal of the Royal Statistical Society, Series {A}},
  147\penalty0 (2):\penalty0 278--292, 1984.

\bibitem[Dziugaite and Roy(2017)]{dziugaite2017computing}
Gintare~Karolina Dziugaite and Daniel~M Roy.
\newblock Computing nonvacuous generalization bounds for deep (stochastic)
  neural networks with many more parameters than training data.
\newblock In \emph{Proceedings of the Thirty-Third Conference on Uncertainty in
  Artificial Intelligence (UAI '17)}, 2017.

\bibitem[Eggeling et~al.(2014)Eggeling, Roos, Myllym{\"{a}}ki, and
  Grosse]{Eggeling14}
Ralf Eggeling, Teemu Roos, Petri Myllym{\"{a}}ki, and Ivo Grosse.
\newblock Robust learning of inhomogeneous {PMMs}.
\newblock In \emph{Proceedings of the Seventeenth International Conference on
  Artificial Intelligence and Statistics, {AISTATS} 2014}, pages 229--237,
  2014.

\bibitem[Eggeling et~al.(2015)Eggeling, Roos, Myllym{\"{a}}ki, and
  Grosse]{Eggeling15}
Ralf Eggeling, Teemu Roos, Petri Myllym{\"{a}}ki, and Ivo Grosse.
\newblock Inferring intra-motif dependencies of {DNA} binding sites from
  chip-seq data.
\newblock \emph{{BMC} Bioinformatics}, 16:\penalty0 375:1--375:15, 2015.

\bibitem[{\VANDER{Erven}{Van}{van}}~Erven
  et~al.(2007){\VANDER{Erven}{Van}{van}}~Erven, Gr\"unwald, and
  de~Rooij]{ErvenGR07}
T.~{\VANDER{Erven}{Van}{van}}~Erven, P.D. Gr\"unwald, and S.~de~Rooij.
\newblock Catching up faster in {B}ayesian model selection and model averaging.
\newblock In \emph{Advances in Neural Information Processing Systems},
  volume~20, 2007.

\bibitem[{\VANDER{Erven}{Van}{van}}~Erven
  et~al.(2012){\VANDER{Erven}{Van}{van}}~Erven, Gr\"unwald, and
  de~Rooij]{erven2012catching}
T.~{\VANDER{Erven}{Van}{van}}~Erven, P.D. Gr\"unwald, and S.~de~Rooij.
\newblock Catching up faster by switching sooner: a predictive approach to
  adaptive estimation with an application to the {AIC--BIC} dilemma.
\newblock \emph{Journal of the Royal Statistical Society: Series B (Statistical
  Methodology)}, 74\penalty0 (3):\penalty0 361--417, 2012.
\newblock With discussion, pp. 399--417.

\bibitem[Geiger and Heckerman(1995)]{Geiger:1995:CDD:2074158.2074181}
Dan Geiger and David Heckerman.
\newblock A characterization of the {D}irichlet distribution with application
  to learning {B}ayesian networks.
\newblock In \emph{Proc. 11th Conference on Uncertainty in Artificial
  Intelligence (UAI-1995)}, pages 196--207, 1995.

\bibitem[Gr\"unwald(2007)]{Grunwald07}
P.~Gr\"unwald.
\newblock \emph{The Minimum Description Length Principle}.
\newblock MIT Press, Cambridge, MA, 2007.

\bibitem[Gr{\"u}nwald(2012)]{Grunwald12}
P.~Gr{\"u}nwald.
\newblock The safe {B}ayesian: learning the learning rate via the mixability
  gap.
\newblock In \emph{Proceedings 23rd International Conference on Algorithmic
  Learning Theory (ALT '12)}. Springer, 2012.

\bibitem[Gr\"unwald and Langford(2007)]{GrunwaldL07}
P.~Gr\"unwald and J.~Langford.
\newblock Suboptimal behavior of {B}ayes and {MDL} in classification under
  misspecification.
\newblock \emph{Machine Learning}, 66\penalty0 (2-3):\penalty0 119--149, 2007.
\newblock DOI 10.1007/s10994-007-0716-7.

\bibitem[Gr{\"u}nwald and Mehta(2019)]{GrunwaldM19}
P.. Gr{\"u}nwald and N.~Mehta.
\newblock A tight excess risk bound via a unified
  {P}{A}{C}-{B}ayesian-{R}ademacher-{S}htarkov-{M}{D}{L} complexity.
\newblock In \emph{Proceedings Thirtieth Conference on Algorithmic Learning
  Theory (ALT 2019)}, 2019.
\newblock A longer version with more links to {MDL} is on arXiv: {\tt
  1720.07732}.

\bibitem[Gr\"unwald and van Ommen(2017)]{GrunwaldO17}
P.~Gr\"unwald and T.~van Ommen.
\newblock Inconsistency of {B}ayesian inference for misspecified linear models,
  and a proposal for repairing it.
\newblock \emph{Bayesian Analysis}, 12\penalty0 (4):\penalty0 1069--1103, 2017.

\bibitem[Gr{\"u}nwald et~al.(2019)Gr{\"u}nwald, de~Heide, and
  Koolen]{GrunwaldHK19}
P.~Gr{\"u}nwald, R.~de~Heide, and W.~Koolen.
\newblock Safe testing, 2019.
\newblock arXiv preprint 1906.07801.

\bibitem[Gr\"unwald(1999)]{Grunwald99a}
P.~D. Gr\"unwald.
\newblock Viewing all models as ``probabilistic''.
\newblock In \emph{Proceedings of the Twelfth ACM Conference on Computational
  Learning Theory (COLT' 99)}, pages 171--182, 1999.

\bibitem[Gr{\"u}nwald(2018)]{grunwald2018safe}
Peter Gr{\"u}nwald.
\newblock Safe probability.
\newblock \emph{Journal of Statistical Planning and Inference}, 195:\penalty0
  47--63, 2018.

\bibitem[Gr{\"u}nwald and Harremo{\"e}s(2009)]{grunwald2009finiteness}
Peter Gr{\"u}nwald and Peter Harremo{\"e}s.
\newblock Finiteness of redundancy, regret, {S}htarkov sums, and {J}effreys
  integrals in exponential families.
\newblock In \emph{Information Theory, 2009. ISIT 2009. IEEE International
  Symposium on}, pages 714--718. IEEE, 2009.

\bibitem[Gr{\"u}nwald and Kot{\l}owski(2010)]{GrunwaldK10}
Peter Gr{\"u}nwald and Wojciech Kot{\l}owski.
\newblock Prequential plug-in codes that achieve optimal redundancy rates even
  if the model is wrong.
\newblock In \emph{Proceedings of the 2010 International Symposium on
  Information Theory (ISIT)}, 2010.

\bibitem[Gr{\"u}nwald et~al.(2005)Gr{\"u}nwald, Myung, and
  Pitt]{grunwald2005advances}
Peter~D Gr{\"u}nwald, In~Jae Myung, and Mark~A Pitt.
\newblock \emph{Advances in minimum description length: Theory and
  applications}.
\newblock MIT press, 2005.

\bibitem[Hastie et~al.(2001)Hastie, Tibshirani, and Friedman]{HastieTF01}
T.~Hastie, R.~Tibshirani, and J.~Friedman.
\newblock \emph{The Elements of Statistical Learning: Data Mining, Inference
  and Prediction}.
\newblock Springer Verlag, 2001.

\bibitem[Heckerman et~al.(1995)Heckerman, Geiger, and
  Chickering]{heckerman1995learning}
David Heckerman, Dan Geiger, and David Chickering.
\newblock Learning {B}ayesian networks: The combination of knowledge and
  statistical data.
\newblock \emph{Machine learning}, 20\penalty0 (3):\penalty0 197--243, 1995.

\bibitem[{\VANDER{Heide}{De}{de}}~Heide(2016)]{de2016safe}
R~{\VANDER{Heide}{De}{de}}~Heide.
\newblock The {S}afe--{B}ayesian {L}asso.
\newblock Master's thesis, Leiden University, 2016.

\bibitem[Hinton and Van~Camp(1993)]{hinton1993keeping}
Geoffrey~E Hinton and Drew Van~Camp.
\newblock Keeping the neural networks simple by minimizing the description
  length of the weights.
\newblock In \emph{Proceedings of the sixth annual conference on Computational
  learning theory}, pages 5--13. ACM, 1993.

\bibitem[Hirai and Yamanishi(2013)]{hirai2013efficient}
So~Hirai and Kenji Yamanishi.
\newblock Efficient computation of normalized maximum likelihood codes for
  gaussian mixture models with its applications to clustering.
\newblock \emph{IEEE Transactions on Information Theory}, 59\penalty0
  (11):\penalty0 7718--7727, 2013.

\bibitem[Hirai and Yamanishi(2017)]{HiraiY17}
So~Hirai and Kenji Yamanishi.
\newblock An upper bound on normalized maximum likelihood codes for gaussian
  mixture models.
\newblock \emph{CoRR}, abs/1709.00925, 2017.
\newblock URL \url{http://arxiv.org/abs/1709.00925}.

\bibitem[Hochreiter and Schmidhuber(1997)]{HochreiterS97}
S.~Hochreiter and J.~Schmidhuber.
\newblock Flat minima.
\newblock \emph{Neural Computation}, 9\penalty0 (1):\penalty0 1--42, 1997.

\bibitem[Kass and Raftery(1995)]{kass1995bayes}
Robert~E Kass and Adrian~E Raftery.
\newblock Bayes factors.
\newblock \emph{Journal of the american statistical association}, 90\penalty0
  (430):\penalty0 773--795, 1995.

\bibitem[Kawakita and Takeuchi(2016)]{KawakitaT16}
Masanori Kawakita and Jun-ichi Takeuchi.
\newblock Barron and {C}over's theory in supervised learning and its
  application to {L}asso.
\newblock In \emph{Proceedings of The 33rd International Conference on Machine
  Learning}, pages 1958--1966, 2016.

\bibitem[Kojima and Komaki(2016)]{kojima2016relations}
Mutsuki Kojima and Fumiyasu Komaki.
\newblock Relations between the conditional normalized maximum likelihood
  distributions and the latent information priors.
\newblock \emph{IEEE Transactions on Information Theory}, 62\penalty0
  (1):\penalty0 539--553, 2016.

\bibitem[Koller and Friedman(2009)]{Koller-book}
Daphne Koller and Nir Friedman.
\newblock \emph{Probabilistic Graphical Models: Principles and Techniques}.
\newblock MIT Press, 2009.

\bibitem[Kontkanen and Myllym{\"a}ki(2007)]{kontkanen2007linear}
Petri Kontkanen and Petri Myllym{\"a}ki.
\newblock A linear-time algorithm for computing the multinomial stochastic
  complexity.
\newblock \emph{Information Processing Letters}, 103\penalty0 (6):\penalty0
  227--233, 2007.

\bibitem[Kot{\l}owski et~al.(2010)Kot{\l}owski, Gr{\"u}nwald, and
  de~Rooij]{kotlowski2010following}
Wojciech Kot{\l}owski, P~Gr{\"u}nwald, and Steven de~Rooij.
\newblock Following the flattened leader.
\newblock In \emph{Conference on Learning Theory (COLT)}, pages 106--118, 2010.

\bibitem[Koutra et~al.(2015)Koutra, Kang, Vreeken, and
  Faloutsos]{koutra2015summarizing}
Danai Koutra, U~Kang, Jilles Vreeken, and Christos Faloutsos.
\newblock Summarizing and understanding large graphs.
\newblock \emph{Statistical Analysis and Data Mining: The ASA Data Science
  Journal}, 8\penalty0 (3):\penalty0 183--202, 2015.

\bibitem[Lam and Bacchus(1994)]{LamBacchus1994}
Wai Lam and Fahiem Bacchus.
\newblock Learning {B}ayesian belief networks: an approach based on the {MDL}
  principle.
\newblock \emph{Computational Intelligence}, 10\penalty0 (3):\penalty0
  269--293, 1994.

\bibitem[Langford and Caruana(2002)]{LangfordC02}
John Langford and Rich Caruana.
\newblock ({N}ot) bounding the true error.
\newblock In T.~G. Dietterich, S.~Becker, and Z.~Ghahramani, editors,
  \emph{Advances in Neural Information Processing Systems 14}, pages 809--816.
  MIT Press, 2002.

\bibitem[Li(1999)]{Li99}
J.Q. Li.
\newblock \emph{Estimation of Mixture Models}.
\newblock PhD thesis, Yale University, New Haven, CT, 1999.

\bibitem[Li and Barron(2000)]{LiB00}
J.Q. Li and A.R. Barron.
\newblock Mixture density estimation.
\newblock In S.A. Solla, T.K. Leen, and K-R. M\"uller, editors, \emph{Advances
  in Neural Information Processing Systems}, volume~12, pages 279--285,
  Cambridge, MA, 2000. MIT Press.

\bibitem[Ly et~al.(2016)Ly, Verhagen, and Wagenmakers]{LiVW16}
A.~Ly, J.~Verhagen, and E.J. Wagenmakers.
\newblock Harold {J}effrey{s'} default {B}ayes factor hypothesis tests:
  Explanation, extension, and application in psychology.
\newblock \emph{Journal of Mathematical Psychology}, 72:\penalty0 19--32, 2016.

\bibitem[M\"a\"att\"a and Roos(2016)]{maatta16b}
Jussi M\"a\"att\"a and Teemu Roos.
\newblock Robust sequential prediction in linear regression with {Student's}
  $t$-distribution.
\newblock In \emph{Proceedings of the Fourteenth International Symposium on
  Artificial Intelligence and Mathematics (ISAIM-2016)}, 2016.

\bibitem[M\"a\"att\"a et~al.(2016)M\"a\"att\"a, Schmidt, and Roos]{maatta16a}
Jussi M\"a\"att\"a, Daniel~F. Schmidt, and Teemu Roos.
\newblock Subset selection in linear regression using sequentially normalized
  least squares: Asymptotic theory.
\newblock \emph{Scandinavian Journal of Statistics}, 43\penalty0 (2):\penalty0
  382--395, 2016.

\bibitem[McAllester(2003)]{McAllester02}
D.~McAllester.
\newblock {PAC}-{B}ayesian stochastic model selection.
\newblock \emph{Machine Learning}, 51\penalty0 (1):\penalty0 5--21, 2003.

\bibitem[Miyaguchi and Yamanishi(2018)]{MiyaguchiY18}
K.~Miyaguchi and K.~Yamanishi.
\newblock High-dimensional penalty selection via minimum description length
  principle.
\newblock \emph{Machine Learning Journal}, 107\penalty0 (8-10):\penalty0
  1283--1302, 2018.

\bibitem[Miyaguchi et~al.(2017)Miyaguchi, Matsushima, and
  Yamanishi]{MiyaguchiMY17}
Kohei Miyaguchi, Shin Matsushima, and Kenji Yamanishi.
\newblock Sparse graphical modeling via stochastic complexity.
\newblock In \emph{Proceedings of 2017 SIAM International Conference on Data
  Mining (SDM2017)}, pages 723--731, 2017.

\bibitem[Miyamoto et~al.(2019)Miyamoto, Barron, and Takeuchi]{MyamotoBT19}
Kohei Miyamoto, Andrew~R. Barron, and J.~Takeuchi.
\newblock Improved mdl estimators using local exponential family bundles
  applied to mixture families.
\newblock In \emph{Proceedings ISIT 2019}, 2019.

\bibitem[Myung et~al.(2000)Myung, Balasubramanian, and Pitt]{MyungBP00}
I.J. Myung, V.~Balasubramanian, and M.A. Pitt.
\newblock Counting probability distributions: Differential geometry and model
  selection.
\newblock \emph{Proceedings of the National Academy of Sciences USA},
  97:\penalty0 11170--11175, 2000.

\bibitem[{\VANDER{Pas}{Van der}{van der}}~Pas and Gr\"unwald(2018)]{PasG17}
S.~{\VANDER{Pas}{Van der}{van der}}~Pas and P~D Gr\"unwald.
\newblock Almost the best of three worlds: Risk, consistency and optional
  stopping for the switch criterion in nested model selection.
\newblock \emph{Statistica Sinica}, 2018.

\bibitem[Rasmussen and Ghahramani(2000)]{RasmussenG00}
C.E. Rasmussen and Z.~Ghahramani.
\newblock Occam's razor.
\newblock In \emph{Advances in Neural Information Processing Systems},
  volume~13, pages 294--300, 2000.

\bibitem[Rissanen(1978)]{Rissanen78}
J.~Rissanen.
\newblock Modeling by the shortest data description.
\newblock \emph{Automatica}, 14:\penalty0 465--471, 1978.

\bibitem[Rissanen(1984)]{Rissanen84}
J.~Rissanen.
\newblock Universal coding, information, prediction and estimation.
\newblock \emph{IEEE Transactions on Information Theory}, 30:\penalty0
  629--636, 1984.

\bibitem[Rissanen(1989)]{Rissanen89}
J.~Rissanen.
\newblock \emph{Stochastic Complexity in Statistical Inquiry}.
\newblock World Scientific, Hackensack, NJ, 1989.

\bibitem[Rissanen(1991)]{Rissanen91}
J.~Rissanen.
\newblock Complexity of models.
\newblock In W.H. Zurek, editor, \emph{Complexity, {E}ntropy and the {P}hysics
  of {I}nformation}, pages 117--125. Addison-Wesley, 1991.

\bibitem[Rissanen(1996)]{Rissanen96}
J.~Rissanen.
\newblock Fisher information and stochastic complexity.
\newblock \emph{IEEE Transactions on Information Theory}, 42\penalty0
  (1):\penalty0 40--47, 1996.

\bibitem[Rissanen(2007)]{Rissanen07}
J.~Rissanen.
\newblock \emph{Information and Complexity in Statistical Modeling}.
\newblock Springer-Verlag, New York, 2007.

\bibitem[Rissanen et~al.(2010)Rissanen, Roos, and {Myllym\"aki}]{RissanenRM10}
J.~Rissanen, T.~Roos, and P.~{Myllym\"aki}.
\newblock Model selection by sequentially normalized least squares.
\newblock \emph{Journal of Multivariate Analysis}, 101\penalty0 (4):\penalty0
  839--849, 2010.

\bibitem[Roos(2008)]{itw2008}
Teemu Roos.
\newblock Monte {C}arlo estimation of minimax regret with an application to
  {MDL} model selection.
\newblock In \emph{Proc. IEEE Information Theory Workshop 2008 (ITW-2008)},
  pages 284 --288. IEEE Press, 2008.

\bibitem[Roos and Rissanen(2008)]{RoosRissanen2008}
Teemu Roos and Jorma Rissanen.
\newblock On sequentially normalized maximum likelihood models.
\newblock In \emph{Proceedings of the First Workshop on Information Theoretic
  Methods in Science and Engineering (WITMSE-2008)}. Tampere International
  Center for Signal Processing, 2008.

\bibitem[Shalev-Shwartz and Ben-David(2014)]{shalev2014understanding}
Shai Shalev-Shwartz and Shai Ben-David.
\newblock \emph{Understanding machine learning: From theory to algorithms}.
\newblock Cambridge university press, 2014.

\bibitem[Shtarkov(1987)]{Shtarkov87}
Yu.~M. Shtarkov.
\newblock Universal sequential coding of single messages.
\newblock \emph{Problems of Information Transmission}, 23\penalty0
  (3):\penalty0 3--17, 1987.

\bibitem[Silander et~al.(2010)Silander, Roos, and Myllym\"aki]{ijar-paper}
T.~Silander, T.~Roos, and P.~Myllym\"aki.
\newblock Learning locally minimax optimal {B}ayesian networks.
\newblock \emph{International Journal of Approximate Reasoning}, 51\penalty0
  (5):\penalty0 544--557, 2010.

\bibitem[Silander et~al.(2018)Silander, Lepp{\"{a}}{-}aho,
  J{\"{a}}{\"{a}}saari, and Roos]{Silander2018}
Tomi Silander, Janne Lepp{\"{a}}{-}aho, Elias J{\"{a}}{\"{a}}saari, and Teemu
  Roos.
\newblock Quotient normalized maximum likelihood criterion for learning
  {B}ayesian network structures.
\newblock In \emph{International Conference on Artificial Intelligence and
  Statistics, {AISTATS} 2018}, pages 948--957, 2018.

\bibitem[Sterkenburg(2018)]{sterkenburg2018universal}
Tom~Florian Sterkenburg.
\newblock \emph{Universal prediction: a philosophical investigation}.
\newblock PhD thesis, University of Groningen, 2018.

\bibitem[Suzuki and Yamanishi(2018)]{suzuki2018exact}
Atsushi Suzuki and Kenji Yamanishi.
\newblock Exact calculation of normalized maximum likelihood code length using
  {F}ourier analysis.
\newblock \emph{arXiv preprint arXiv:1801.03705}, 2018.

\bibitem[Suzuki et~al.(2016)Suzuki, Miyaguchi, and
  Yamanishi]{suzuki2016structure}
Atsushi Suzuki, Kohei Miyaguchi, and Kenji Yamanishi.
\newblock Structure selection for convolutive non-negative matrix factorization
  using normalized maximum likelihood coding.
\newblock In \emph{2016 IEEE 16th International Conference on Data Mining
  (ICDM)}, pages 1221--1226. IEEE, 2016.

\bibitem[Suzuki(2016)]{profJoe-witmse2016}
Joe Suzuki.
\newblock Jeffreys' and {BDeu} priors for model selection.
\newblock In J.~Rissanen, J.~Lepp\"a-aho, T.~Roos, and P.~Myllym\"aki, editors,
  \emph{Proc. 9th Workshop on Information Theoretic Methods in Science and
  Engineering (WITMSE-2016)}, 2016.

\bibitem[Szpankowski and Weinberger(2012)]{Szpankowski2012}
Wojciech Szpankowski and Marcelo~J. Weinberger.
\newblock Minimax pointwise redundancy for memoryless models over large
  alphabets.
\newblock \emph{{IEEE} Trans. Information Theory}, 58\penalty0 (7):\penalty0
  4094--4104, 2012.

\bibitem[Takeuchi and Barron(1998)]{TakeuchiB98b}
J.~Takeuchi and A.~R. Barron.
\newblock Robustly minimax codes for universal data compression.
\newblock In \emph{Proceedings of the Twenty-First Symposium on Information
  Theory and Its Applications (SITA '98)}, Gifu, Japan, 1998.

\bibitem[Takimoto and Warmuth(2000)]{takimoto00}
Eiji Takimoto and Manfred Warmuth.
\newblock The last-step minimax algorithm.
\newblock In \emph{Proceedings of the Eleventh International Conference on
  Algorithmic Learning Theory (ALT-2000)}, 2000.

\bibitem[Tatti and Vreeken(2008)]{tatti2008finding}
Nikolaj Tatti and Jilles Vreeken.
\newblock Finding good itemsets by packing data.
\newblock In \emph{Data Mining, 2008. ICDM'08. Eighth IEEE International
  Conference on}, pages 588--597. IEEE, 2008.

\bibitem[Vit\'anyi and Li(2000)]{VitanyiL00}
P.M.B. Vit\'anyi and M.~Li.
\newblock Minimum description length induction, {B}ayesianism, and {K}olmogorov
  complexity.
\newblock \emph{IEEE Transactions on Information Theory}, IT-46\penalty0
  (2):\penalty0 446--464, 2000.

\bibitem[Vreeken et~al.(2011)Vreeken, van Leeuwen, and Siebes]{VreekenLS11}
Jilles Vreeken, Matthijs van Leeuwen, and Arno Siebes.
\newblock Krimp: mining itemsets that compress.
\newblock \emph{Dat Mining and Knowledge Discovery}, 23\penalty0 (1):\penalty0
  169--214, 2011.

\bibitem[Wallace and Boulton(1968)]{WallaceB68}
C.S. Wallace and D.M. Boulton.
\newblock An information measure for classification.
\newblock \emph{Computer Journal}, 11:\penalty0 185--195, 1968.

\bibitem[Watanabe and Roos(2015)]{watanabe2015achievability}
Kazuho Watanabe and Teemu Roos.
\newblock Achievability of asymptotic minimax regret by horizon-dependent and
  horizon-independent strategies.
\newblock \emph{Journal of Machine Learning Research}, 16\penalty0
  (23572375):\penalty0 11, 2015.

\bibitem[Watanabe(2010)]{watanabe2010}
Sumio Watanabe.
\newblock Asymptotic equivalence of {Bayes} cross validation and widely
  applicable information criterion in singular learning theory.
\newblock \emph{Journal of Machine Learning Research}, 11:\penalty0 3571--3594,
  Dec 2010.

\bibitem[Watanabe(2013)]{watanabe2013}
Sumio Watanabe.
\newblock A widely applicable {Bayesian} information criterion.
\newblock \emph{Journal of Machine Learning Research}, 14:\penalty0 867--897,
  Mar 2013.

\bibitem[Wu et~al.(2017)Wu, Sugawara, and Yamanishi]{WuSY17}
Tianyi Wu, Shinya Sugawara, and Kenji Yamanishi.
\newblock Decomposed normalized maximum likelihood codelength criterion for
  selecting hierarchical latent variable models.
\newblock In \emph{ACM International Conference on Knowledge Discovery and Data
  Mining}, 2017.

\bibitem[Yang(2005)]{Yang05a}
Y.~Yang.
\newblock Can the strengths of {AIC} and {BIC} be shared? {A} conflict between
  model indentification and regression estimation.
\newblock \emph{Biometrica}, 92\penalty0 (4):\penalty0 937--950, 2005.

\bibitem[Zhang(2006{\natexlab{a}})]{Zhang06a}
Tong Zhang.
\newblock From $\epsilon$-entropy to {KL} entropy: analysis of minimum
  information complexity density estimation.
\newblock \emph{Annals of Statistics}, 34\penalty0 (5):\penalty0 2180--2210,
  2006{\natexlab{a}}.

\bibitem[Zhang(2006{\natexlab{b}})]{Zhang06b}
Tong Zhang.
\newblock Information theoretical upper and lower bounds for statistical
  estimation.
\newblock \emph{IEEE Transactions on Information Theory}, 52\penalty0
  (4):\penalty0 1307--1321, 2006{\natexlab{b}}.

\bibitem[Zhou et~al.(2018)Zhou, Veitch, Austern, Adams, and
  Orbanz]{zhou2018compressibility}
Wenda Zhou, Victor Veitch, Morgane Austern, Ryan Adams, and Peter Orbanz.
\newblock Compressibility and generalization in large-scale deep learning.
\newblock \emph{arXiv preprint arXiv:1804.05862}, 2018.

\bibitem[Zou and Roos(2017)]{ngc2017}
Yuan Zou and Teemu Roos.
\newblock On model selection, {B}ayesian networks, and the {F}isher information
  integral.
\newblock \emph{New Generation Computing}, 35\penalty0 (1):\penalty0 5--27,
  2017.

\end{thebibliography}
\appendix
\section{When the original ($v \equiv 1$)  NML is undefined: Details, Open Problems and their Solutions}\label{app:nml}
The original NML distribution $\ud{\nml}{}$ with uniform $v$ relies on the existence of the Shtarkov
integral $\int_{z^n \in \cZ^n} p_{\mlest{\theta}(z^n)}(z^n) dz^n$; its
asymptotic expansion (\ref{eq:nmlasymptotics}) relies on the existence of the {\em Jeffreys integral\/} $\int \sqrt{|I(\theta)} d \theta$ being
finite, the latter being equivalent to the requirement that Jeffreys'
prior is proper. Both are quite strong requirements; for infinite
sample spaces $\cY$, they `usually' --- that is, in most models one
considers in practice, such as normal, exponential, Poisson ... --- do
not hold; but once one restricts the parameter space to an INECCSI
set, they generally do hold. This may lead one to conjecture that the
Shtarkov integral is finite {\em if and only if \/} the corresponding
Jeffreys integral is finite. Resolving this conjecture was posed as an
open problem by G07;
\cite{grunwald2009finiteness,bar2010jeffreys} show that in general,
the conjecture is wrong; though, for exponential families, under a
very mild additional  condition, it holds true.

From a more practical perspective, one would of course like to know
what universal distribution to use if the standard MDL is
undefined. Several proposals floated around in the early 2000s; for an
overview, see Chapter 11 of G07. By now, the dominant
method has become to factor in a nonuniform weight function $v$ and
calculate the {\em luckiness NML\/} as in (\ref{eq:expregret}). This
method was originally called {\em luckiness NML-2\/} by
G07, who (among many other methods) identified several
'luckiness' versions of NML that had been proposed by various authors; luckiness NML-2 turned out both more practically useable and
mathematically analyzable than other methods, and in this text we simply call it {\em luckiness NML}. In particular, \cite{suzuki2018exact} show that, for exponential family models, the $n$-dimensional integral in the luckiness NML can be replaced by a $2k$-dimensional one, and in many cases can be performed explicitly. As we indicated in Section~\ref{sec:luckiness}, one can sometimes set the first $m$ examples aside as start-up data to define a luckiness function, leading to {\em conditional NML}. 
Again, G07 defined different forms of conditional NML, and again, conditional NML-2 (directly based on luckiness NML-2) turned out to be the most natural one: 
\cite{bartlett2013horizon} show that for some important classes of models, the NML distributions $\ud{\nml}{}$ and the Bayes marginal
distributions $\ud{\bayes}{}$ with improper Jeffreys' prior {\em exactly}, and not just asymptotically,
coincide for each $n$. Moreover, for the case of 1-dimensional families, they completely characterize the
class of models for which this holds: essentially,
it holds for exponential families that are also  location
or scale families, i.e. the normal and gamma distributions, and monotone transformations thereof (such as
e.g. the Rayleigh distributions); as well as for one curious
additional family. This correspondence between objective Bayesian and conditional NML-2 approaches notwithstanding,
\cite{kojima2016relations} show that `conditional NML-3', which G07 considered the most intuitive version, but at the same time, mathematically overly complicated for practical use,  can be given a practical implementation after all, thereby solving Open Problem 7 of G07.

\end{document}